\documentclass[twocolumn,showpacs,preprintnumbers,amsmath,amssymb,superscriptaddress,prb,dvipdfmx]{revtex4-1}
\usepackage[dvipdfmx]{graphicx}
\usepackage{dcolumn}
\usepackage{bm}
\usepackage{amsmath}	
\usepackage{txfonts,wasysym,ulem,url}
\usepackage[usenames]{color}

\newcommand{\red}[1]{#1}

\newcommand{\rmd}{{\rm d}}
\renewcommand{\vec}[1]{{\bm #1}}

\begin{document}

\title{Self-consistent van der Waals density functional study of benzene adsorption on Si(100)}

\author{Yuji Hamamoto}
\email{hamamoto@prec.eng.osaka-u.ac.jp}
\affiliation{Department of Precision Science and Technology, Graduate School of Engineering,
Osaka University, Suita, Osaka 565-0871, Japan}

\affiliation{ACT-C, Japan Science and Technology Agency (JST),
Kawaguchi, Saitama 332-0012, Japan}

\affiliation{Elements Strategy Initiative for Catalysts and Batteries (ESICB),
Kyoto University, Katsura, Kyoto 615-8520, Japan}

\author{Ikutaro Hamada}
\affiliation{International Research Center for Materials Nanoarchitectonics (WPI-MANA)
and Global Research Center for Environment and Energy based on Nanomaterials Science (GREEN),
National Institute for Materials Science (NIMS), 1-1 Namiki, Tsukuba 305-0044, Japan}

\author{Kouji Inagaki}
\affiliation{Department of Precision Science and Technology, Graduate School of Engineering,
Osaka University, Suita, Osaka 565-0871, Japan}

\affiliation{ACT-C, Japan Science and Technology Agency (JST),
Kawaguchi, Saitama 332-0012, Japan}

\affiliation{Elements Strategy Initiative for Catalysts and Batteries (ESICB),
Kyoto University, Katsura, Kyoto 615-8520, Japan}

\author{Yoshitada Morikawa}
\affiliation{Department of Precision Science and Technology, Graduate School of Engineering,
Osaka University, Suita, Osaka 565-0871, Japan}

\affiliation{ACT-C, Japan Science and Technology Agency (JST),
Kawaguchi, Saitama 332-0012, Japan}

\affiliation{Elements Strategy Initiative for Catalysts and Batteries (ESICB),
Kyoto University, Katsura, Kyoto 615-8520, Japan}

\affiliation{Research Center for Ultra-Precision Science and Technology, Graduate School of Engineering,
Osaka University, Suita, Osaka 565-0871, Japan}

\begin{abstract}
The adsorption of benzene on the Si(100) surface is studied theoretically
using the self-consistent van der Waals density functional (vdW-DF) method.
The adsorption energies of two competing adsorption structures, butterfly (BF)
and tight-bridge (TB) structures, are calculated with several vdW-DFs at saturation coverage.
\red{Our results show that recently proposed vdW-DFs with high accuracy all prefer TB to BF,
in accord with more accurate calculations based on exact exchange and correlation
within the random phase approximation. }
\red{Detailed analyses reveal the important roles played by the molecule-surface interaction
and molecular deformation upon adsorption, and we suggest that their precise description 
is prerequisite for accurate prediction of the most stable adsorption structure of organic molecules 
on semiconductor surfaces.}
\end{abstract}


\maketitle

\section{Introduction}
\begin{figure}[b]
\hfill
\includegraphics[width=.45\linewidth]{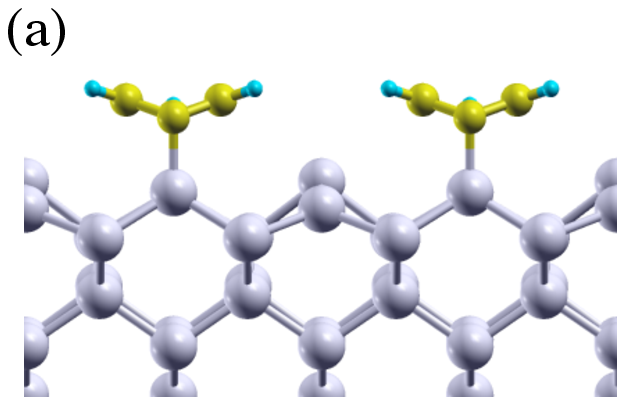}
\hfill
\includegraphics[width=.45\linewidth]{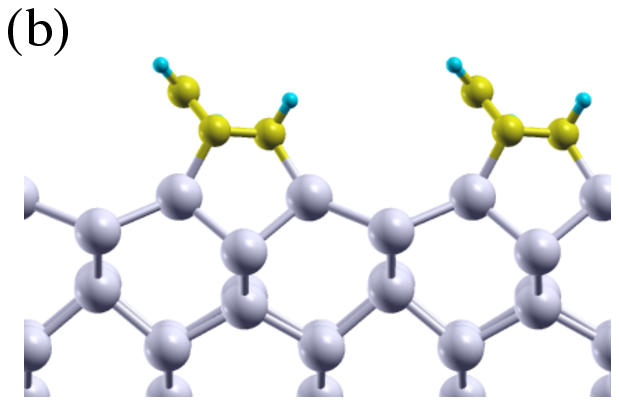}
\hfill{}
\caption{\label{fig:structure}(Color online)
Schematics of the competing adsorption structures of benzene on the Si(100) surface:
(a) butterfly (BF) and (b) tight-bridge (TB) structures.
The side views at 0.5 ML coverage are shown.
}
\end{figure}

The adsorption of benzene on silicon surfaces is one of the best studied subjects
in surface science, since the system plays a prototypical role
in molecular modification of semiconductor surfaces.
In particular, benzene adsorption on the Si(100) surface is a long-disputed problem
due to the lack of decisive evidence of the adsorption structure of benzene.
Several adsorption structures of benzene have been proposed so far,
and it is now widely believed that the most stable structure is either butterfly (BF)
or tight-bridged (TB) structures depicted in Fig.~\ref{fig:structure}.
In the former structure benzene is di-$\sigma$ bonded to a single Si dimer,
while in the latter it is tetra-$\sigma$ bonded to two adjacent dimers.

To determine the most stable adsorption structure,
a wide range of experimental techniques have been
applied to benzene on Si(100),
\cite{JChemPhys.95.1, ChemPhysLett.282.305, JVacSciTechnolA.16.1031,
JVacSciTechnolA.16.1037, PhysRevB.57.R4269, JChemPhys.108.5554, SurfSci.411.286,
SurfSci.479.69, SurfSci.513.413, PhysRevB.68.115408, SurfSci.547.324, SurfRevLett.10.499,
PhysRevB.71.115311, PhysRevB.72.075354, JPhysCondMatt.20.304206, JChemPhys.127.154711}
and most of them have concluded that BF is more preferable than TB.
Scanning tunneling microscopy (STM) studies,
on the other hand, show that BF is metastable
and converts to a bridging configuration on a time scale of minutes at room temperature.
\cite{ChemPhysLett.282.305, JVacSciTechnolA.16.1037, PhysRevB.57.R4269, SurfSci.547.324}
However, the STM measurements of the other groups observe no such conversion,
\cite{JVacSciTechnolA.16.1031, PhysRevB.72.075354}
which is also supported by recent experiments.
\cite{SurfSci.513.413, PhysRevB.68.115408, SurfRevLett.10.499}
Regarding the discrepancy between these experimental results,
an important suggestion
has been provided by photoelectron spectroscopy,
\cite{PhysRevB.71.115311}
which demonstrates that TB is predominant at low coverage,
while BF becomes the major adsorption structure with increasing coverage.
Photoelectron diffraction results also indicate that
the two adsorption configurations indeed coexist
at saturation coverage and room temperature.
\cite{JPhysCondMatt.20.304206}

Along with these experiments, theoretical investigation of benzene on Si(100)
has also been carried out intensively.
\cite{SurfSciLett.280.L279, SurfSci.344.L1226, ChemPhysLett.282.305, SurfSci.409.213,
JVacSciTechnolA.16.1037, JChemPhys.108.5554, SurfSci.416.L1107, SurfSci.417.169, 
PhysRevB.62.1596, SurfSci.482.1181, PhysRevB.63.085314, JAmChemSoc.127.3131, 
PhysRevB.72.235317, PhysRevB.73.035321, PhysRevB.76.085402, PhysRevB.77.121404,
PhysRevB.85.041403, SurfSci.621.152}
While semiempirical cluster calculations favor
unstable adsorption structures such as symmetric- and tilted-bridge configurations,
\cite{SurfSciLett.280.L279, SurfSci.344.L1226,
ChemPhysLett.282.305, SurfSci.409.213, JVacSciTechnolA.16.1037, SurfSci.416.L1107}
density functional theory (DFT) calculations within the local density approximation (LDA)
and the generalized gradient approximation (GGA) predict that TB is
the most stable adsorption structure,
\cite{SurfSci.416.L1107, PhysRevB.62.1596, SurfSci.482.1181, PhysRevB.63.085314,
PhysRevB.72.235317, PhysRevB.73.035321, PhysRevB.76.085402}
in good agreement with the STM results.
\cite{ChemPhysLett.282.305, JVacSciTechnolA.16.1037, PhysRevB.57.R4269, SurfSci.547.324}
On the other hand, BF is supported
\cite{JAmChemSoc.127.3131, PhysRevB.77.121404}
only by the cluster calculation
within the M{\o}ller-Plesset second-order (MP2) perturbation theory
\cite{PhysRev.46.618}
and the periodic DFT calculation with the van der Waals density functional (vdW-DF) method,
\cite{PhysRevLett.92.246401}
which take into account the vdW interaction unlike the conventional semilocal DFT calculations.
Note, however, that the influence of the vdW interaction in the present problem
is yet to be fully understood,
since the cluster model used in the MP2 calculation
\cite{JAmChemSoc.127.3131}
corresponds to the low coverage limit,
while the vdW-DF results
\cite{PhysRevB.77.121404}
show that BF becomes slightly more stable than TB at almost saturation coverage.
Indeed, other DFT calculations with semiempirical dispersion correction
\cite{SurfSci.621.152}
and a more sophisticated method based on exact exchange and
correlation within the random phase approximation (EX + cRPA)
\cite{PhysRevB.85.041403}
support the TB structure,
suggesting that the stability of the two adsorption structures
cannot be inverted only by the vdW interaction.
Moreover, the above vdW-DF study
\cite{PhysRevB.77.121404}
leaves some ambiguities in its accuracy in retrospect.
Namely, it has been well-recognized that the vdW-DF used in this calculation
tends to overestimate equilibrium separations.
\cite{PhysRevLett.92.246401, JPhysCondMat.21.084203, JChemPhys.132.134703}
In addition, the vdW-DF study calculates non-local (NL) correlation energy
non-self-consistently using the
\red{charge density} and geometries determined within GGA.
Although these ambiguities are expected to make only a little difference,
still they cannot be ignored because the vdW-DF study predicts that
the energy difference between BF and TB is as small as 0.05 eV.
\cite{PhysRevB.77.121404}

Recently, there have been rapid progresses in the vdW-DF method
especially in terms of efficiency
\cite{PhysRevLett.103.096102, PhysRevB.79.201105, JChemPhys.136.224107}
and accuracy.
\cite{PhysRevB.81.161104, PhysRevB.82.081101, JPhysCondMatt.22.022201, PhysRevB.83.195131,
TopCatal.54.1143, PhysRevB.85.235149, PhysRevB.89.035412, PhysRevB.89.121103}
They enable one to calculate vdW interaction with higher accuracy,
so that it is highly worth reconsidering the problem of benzene on Si(100)
taking full advantage of these techniques.
In this paper, we theoretically investigate the adsorption structure
of benzene on Si(100) using the self-consistent (SC)
vdW-DF method based on several vdW-DFs.
Our results show that the adsorption energies of BF and TB are quite sensitive to
the choice of vdW-DF, and
in particular, some of the vdW-DFs predict that TB is more stable than BF,
in good agreement with accurate EX-cRPA calculations.
\cite{PhysRevB.85.041403}
A more detailed analysis reveals that the importance of the SC treatment of vdW-DFs
becomes prominent in interaction between benzene and the Si surface
as well as deformation of a benzene molecule.

\section{Methods}
The DFT calculation in the present paper is carried out using the \textsc{State}
\cite{ApplSurfSci.169.11}
code with norm-conserving pseudopotentials.
\cite{PhysRevB.43.1993}
The plane-wave basis set is used with an energy cutoff of 64~Ry (400~Ry)
for wave functions (charge density).
The Si(100) surface is modeled with a periodically repeated slab composed of nine
Si atomic layers.
Benzene is adsorbed on one side of the slab in the BF or TB configuration,
and the other side is passivated with two H atoms per Si atom.
To avoid long-range vdW interaction between the slabs
we use a vacuum layer ($\gtrsim$ 17 {\AA}).
Moreover, artificial electrostatic interaction between the slabs
is corrected by introducing an effective screening medium.
\cite{PhysRevB.73.115407, PhysRevB.80.165411}
Since the photoelectron measurements
\cite{PhysRevB.71.115311, JPhysCondMatt.20.304206}
observe the increase in the ratio of BF at almost saturation coverage,
we here focus on a $2\times2$ unit cell of the Si(100) surface,
which coincides with the 0.5 ML coverage of benzene.
Correspondingly, $4\times4\times1$ $k$-points are sampled in the Brillouin zone.
We relax the whole system except for the two lowest Si layers and the bottom H atoms
using each vdW-DF until the atomic forces fall below
$5.14\times10^{-2}$~eV/{\AA} ($10^{-3}$ Hartree/Bohr).
The Si atoms in the two lowest layers are fixed at the bulk positions
\red{with a lattice constant of 5.47~{\AA} obtained with the Perdew-Burke-Ernzerhof (PBE) functional,
\cite{PhysRevLett.77.3865}}
while the H atoms at the bottom are fixed at positions optimized with PBE
on a fixed Si(100)-($1\times1$) surface.
We have confirmed that the adsorption energy changes by at most 30~meV
even if we adopt the lattice constant optimized for each vdW-DF.

The vdW interaction is taken into account in the framework of DFT
based on the vdW-DF method,
\cite{PhysRevLett.92.246401}
whose exchange-correlation energy takes the form of
\begin{gather}
 E_{\rm xc}^{\rm vdW}=E_{\rm x}^{\rm GGA}+E_{\rm c}^{\rm LDA}+E_{\rm c}^{\rm NL}
\end{gather}
with the GGA exchange energy $E_{\rm x}^{\rm GGA}$ and 
the LDA correlation energy $E_{\rm c}^{\rm LDA}$.
The NL correlation energy
\begin{gather}
 E_{\rm c}^{\rm NL}=\frac{1}{2}\int\rmd\vec{r}\rmd\vec{r}'
n(\vec{r})\phi(d,d')n(\vec{r}')\label{eq:non-local}
\end{gather}
describes long-range interactions through the vdW kernel $\phi$,
which is proportional to
$\propto R^{-6}$ for large spatial distance $R\equiv|\vec{r}-\vec{r}'|$.
At finite distance, on the other hand, $\phi$ is a function of dimensionless distances
$d\equiv q_0(\vec{r})R$ and $d'\equiv q_0(\vec{r}')R$,
where scaling factor $q_0(\vec{r})$ depends on the charge density $n(\vec{r})$
and its reduced gradient $s\equiv|\nabla n|/2k_Fn$
with $k_{\rm F}$ being the Fermi wave number.
To suppress the high computational cost $\sim\mathcal{O}(N^2)$
required for the double spatial integral in Eq.~(\ref{eq:non-local}),
Rom\'an-P\'erez and Soler\cite{PhysRevLett.103.096102} (RPS) have
represented the vdW kernel as a bilinear combination of cubic splines
$\{p_1,p_2,\cdots,p_{N_q}\}$ as
\begin{gather}
 \phi(d,d')\simeq\sum_{\alpha=1}^{N_q}\sum_{\beta=1}^{N_q}\phi_{\alpha\beta}(R)
p_\alpha\bigl(q_0(\vec{r})\bigr)p_\beta\bigl(q_0(\vec{r}')\bigr),\label{eq:rps}
\end{gather}
Here the spline curves satisfy $p_\alpha(q_\beta)=\delta_{\alpha\beta}$
on $q$-mesh points $\{q_1,q_2,\cdots,q_{N_q}\}$,
and the coefficients are defined as $\phi_{\alpha\beta}(R)\equiv\phi(q_\alpha R,q_\beta R)$.
Since $\phi_{\alpha\beta}$ is diagonal with respect to wave numbers in the Fourier space,
the computational cost of $E_{\rm c}^{\rm NL}$ can be reduced to
$\sim\mathcal{O}(N\log N)$ that is required for fast Fourier transform.
This enables efficient vdW-DF calculation.
Note, however, that the approximation~(\ref{eq:rps}) fails at $d,d'\rightarrow0$,
in which $\phi$ diverges logarithmically.
In the original RPS algorithm, the problem is avoided by replacing $\phi$ with a soft kernel
for small $d$ and $d'$, whose resulting error is corrected within LDA.
\cite{PhysRevLett.103.096102}
Wu and Gygi (WG) have introduced a simplified implementation,
where the divergence is suppressed as $dd'\phi(d,d')$, to which the expansion~(\ref{eq:rps})
is applied. Then the vdW kernel is approximated as
\begin{gather}
 \phi(d,d')\simeq\sum_{\alpha\beta}\phi_{\alpha\beta}(R)
\frac{q_\alpha p_\alpha\bigl(q_0(\vec{r})\bigr)}{q_0(\vec{r}')}
\frac{q_\beta p_\beta\bigl(q_0(\vec{r}')\bigr)}{q_0(\vec{r}')}.\label{eq:wg}
\end{gather}
In the latter formulation $\phi$ can be expanded with a finite number of cubic splines
even for small $d$ and $d'$, which leads to further reduction of computational cost.
Thus, we implement the SC vdW-DF method in the \textsc{State} code
using the WG formulation for the RPS algorithm.
The performance of the approximation~(\ref{eq:wg}) will be discussed
in Sec.~\ref{subsec:performance}.

\begin{figure*}[t]
\includegraphics[width=.9\linewidth]{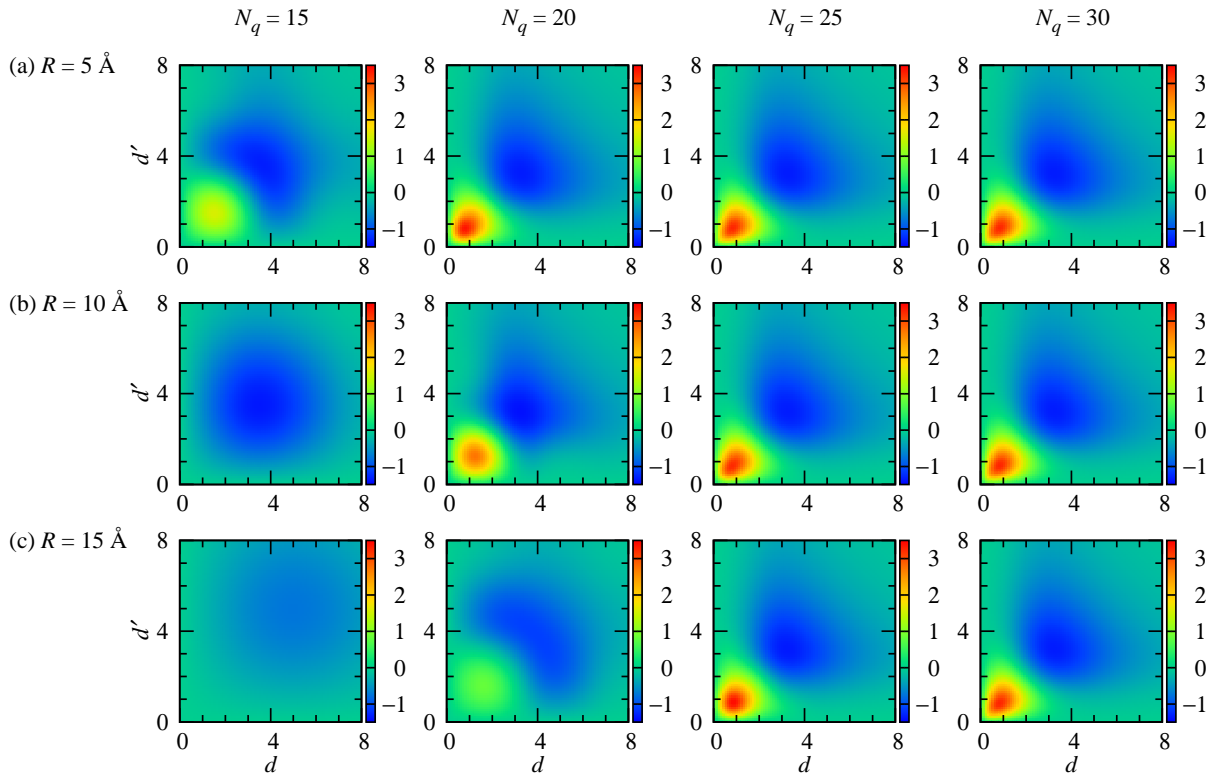}
 \caption{\label{fig:kernel}(Color online)
 vdW kernel approximated in the WG
\red{formulation}~(\ref{eq:wg}).
In each panel $dd'\phi^{\rm WG}$ in Eq.~(\ref{eq:ddphi}) is plotted as a function of 
dimensionless distances $d$ and $d'$ in eV.
The results for spatial distances (a) $R=5$ {\AA}, (b) 10 {\AA} and (c) 15 {\AA} are shown.
The peak at $d=d'\simeq0.8$ decays with increasing $R$ for $N_q=15$ and 20,
while the whole structure of $dd'\phi^{\rm WG}$ is almost independent of $R$
for $N_q=25$ and 30.}
\end{figure*}

In addition to the efficiency of the vdW-DF method,
the accuracy of the vdW-DF has also been improved in the last decade.
As has often been pointed out, the original version of the vdW-DF (vdW-DF1)
\cite{PhysRevLett.92.246401}
systematically overestimates equilibrium separations.
\cite{PhysRevLett.92.246401, JPhysCondMat.21.084203, JChemPhys.132.134703}
This is because the enhancement factor $F_{\rm x}$ of the revised
Perdew-Burke-Ernzerhof (revPBE) exchange
\cite{PhysRevLett.80.890}
adopted in vdW-DF1 rises steeply for small density gradient $s$,
resulting in too large exchange respulsion at a short distance.
In addition, $F_{\rm x}$ of revPBE saturates at large $s$,
giving rise to spurius binding from exchange only.
It has been shown
\cite{PhysRevA.47.4681, JChemTheoryComput.5.719, JChemTheoryComput.5.2754}
that the latter can be avoided by choosing the exchange functional with $F_{\rm x}$
proportional to $s^{2/5}$ at large $s$.
\cite{PhysRevB.33.8800, JChemPhys.85.7184}
To overcome the drawbacks of vdW-DF1,
a variety of exchange functionals have been proposed for the vdW-DF
method\cite{PhysRevB.81.161104, PhysRevB.82.081101, JPhysCondMatt.22.022201,
PhysRevB.83.195131, TopCatal.54.1143, PhysRevB.85.235149, PhysRevB.89.035412,
PhysRevB.89.121103}
and have shown better performances for the benchmark S22 dataset of
non-covalently interacting molecules.\cite{PhysChemChemPhys.8.1985}
In addition to the exchange, the NL correlation part has also been improved
in the second version of the vdW-DF (vdW-DF2),\cite{PhysRevB.82.081101}
where the gradient correction in $q_0(\vec{r})$ is modified so that
it is more suited to atoms and small molecules.
However vdW-DF2 also tends to overestimate separations,
since the $F_{\rm x}$ of the Perdew-Wang exchange with refit parameters
(PW86R)\cite{JChemTheoryComput.5.2754} adopted in vdW-DF2
steeply rises except for sufficiently small $s$.
The overestimation has been avoided in the revised vdW-DF2
(rev-vdW-DF2),\cite{PhysRevB.89.121103}
where the PW86R exchange is replace by the Becke exchange (B86b)\cite{JChemPhys.85.7184}
with revised parameters (B86R).
Recently, rev-vdW-DF2 has been successfully applied to various adsorption systems
\cite{PhysRevB.90.075414, ChemPhysChem.15.2851, ACSNano.8.9181,PhysRevLett.115.236101,JAmChemSoc.137.15}
as well as rare gas and small molecules.\cite{PhysRevB.91.195103}
In the present paper, we use
vdW-DF1, vdW-DF2, opt-vdW-DFs (optPBE-vdW,\cite{JPhysCondMatt.22.022201} optB88-vdW\cite{JPhysCondMatt.22.022201} and optB86b-vdW\cite{PhysRevB.83.195131}) and rev-vdW-DF2
to discuss how the difference among the vdW-DFs influences the relative stability
of the adsorption structures of benzene on Si(100).

\section{Results and discussion}
\subsection{Performance test of the WG formulation}\label{subsec:performance}
The WG formulation improves the accuracy of the kernel decomposition
for small $d$ and $d'$, but the accuracy can decline at large distance $R$,
since the peak of cubic spline $p_\alpha$ scales as $q_\alpha R$ on $d$ and $d'$ axes.
Thus, prior to the application of the SC vdW-DF method to benzene on Si(100),
we here examine the performance of the WG formulation.
To this end, we plot in Fig.~\ref{fig:kernel} the following quantity
\begin{gather}
 dd'\phi^{\rm WG}(d,d',R)\equiv\sum_{\alpha=1}^{N_q}\sum_{\beta=1}^{N_q}
 q_\alpha Rq_\beta R\phi_{\alpha\beta}(R)
p_\alpha(d/R)p_\beta(d'/R)\label{eq:ddphi}
\end{gather}
as a function of $d$ and $d'$ for several values of $R$ and $N_q$.
Here we use a logarithmic mesh such that
$(q_{\alpha+1}-q_\alpha)=\lambda(q_\alpha-q_{\alpha-1})$ with $\lambda=1.2$
whose end points are fixed at $q_1=1.89\times10^{-7}~\AA^{-1}$ ($10^{-7}$~Bohr$^{-1}$)
and $q_{N_q}=18.9~\AA^{-1}$ ($10$~Bohr$^{-1}$).
At $R=5$ \AA, the results for $N_q=20, 25$ and 30 show
a peak $\simeq3.3$~eV at $d=d'\simeq 0.8$ and a dip $\simeq-1.3$~eV at $d=d'\simeq 3.3$.
For $N_q=15$, on the other hand,
the peak is shifted to $d=d'\simeq 1.5$ and its height is reduced to  1.6~eV.
At $R=10$ \AA, the results for $N_q=25$ and 30 remain almost unchanged,
whereas the peak shows a deformation even for $N_q=20$ and completely disappears for $N_q=15$.
At $R=15$ {\AA}, the peak height is further reduced for $N_q=20$
and a slight deformation of the peak can be seen for $N_q=25$ and 30.
The disappearance of the peak is attributed to the lack in cubic splines
that contribute to the small $d$ and $d'$ region,
and can lead to underestimation of vdW interaction at long distance.
Although this suggests that one should use at least $N_q=25$ for physisorbed systems,
benzene adsorption on Si(100) is dominated by covalent bonds at short distance.
Thus we use $N_q=20$ in what follows
and have confirmed that larger $N_q$ changes the adsorption energies only by $\simeq 1$~meV.
We note that the behavior of the vdW kernel in the Fourier space
is discussed in Ref.~\citenum{JPhysSocJpn.84.024715}.

\begin{figure}[t]
 \includegraphics[width=.9\linewidth]{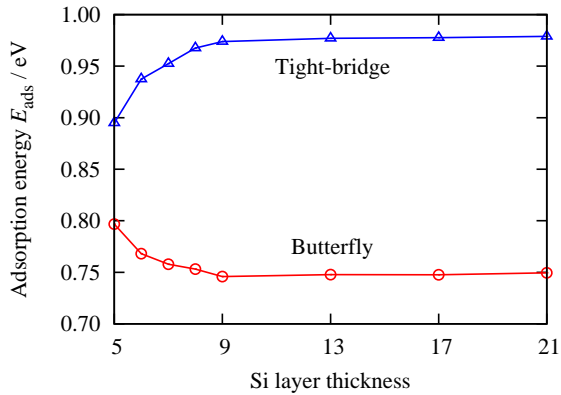}
\caption{\label{fig:thickness}(Color online)
Adsorption energy calculated with PBE as a function of Si layer thickness.
Circles (triangles) show the results for the BF (TB) structure.
}
\end{figure}

\subsection{Adsorption energy of benzene on the Si(100) surface}
We next investigate benzene adsorption on Si(100) using the SC vdW-DF method.
To compare the stabilities of BF and TB, we calculate the adsorption energy defined as
\begin{gather}
 E_{\rm ads}=E_{\rm C_6H_6}+E_{\rm Si}-E_{\rm C_6H_6/Si},\label{eq:energy_ads}
\end{gather}
where $E_{\rm C_6H_6},E_{\rm Si}$ and $E_{\rm C_6H_6/Si}$
are the total energies of an isolated benzene molecule,
a clean Si surface and the adsorbed system, respectively.
To model the clean Si surface,
we use a $4\times2$ unit cell of the Si(100) surface with asymmetric Si dimers.
In Fig.~\ref{fig:thickness}, $E_{\rm ads}$ obtained
\red{with} PBE is plotted
as a function of Si layer thickness.
$E_{\rm ads}$(BF) [$E_{\rm ads}$(TB)] decreases (increases) monotonically
from five to nine layers, whereas the thickness dependence is almost negligible
for more than nine layers.
Thus we here focus on the slab model with nine Si layers
to calculate $E_{\rm ads}$ using the SC vdW-DF method.

The results of $E_{\rm ads}$ for each adsorption structure and vdW-DF
are summarized in Table~\ref{tbl:ads_energy},
where PBE results are also shown for comparison.
In the full vdW-DF method denoted by ``SC-relaxed,''
both SC calculation and structure relaxation are carried out using each vdW-DF.
On the other hand, ``SC-fixed'' indicates that SC calculation is performed with each vdW-DF,
while the structure is fixed to the PBE geometry.
In the ``non-SC'' vdW-DF method, total energy is calculated in a post-processing manner
using the charge density and geometry obtained with PBE.
From the table one readily notices that $E_{\rm ads}$ differs only slightly
among the SC-relaxed, SC-fixed and non-SC results,
in analogy with binding energies of noble gases and small molecules.
\cite{PhysRevB.76.125112}
Of more importance is the fact that $E_{\rm ads}$ is strongly dependent on
the choice of vdW-DF.
For example, the vdW-DF1 and vdW-DF2 results differ from the PBE ones
by less than $\pm0.1$ eV,
while those obtained 
\red{with} the other vdW-DFs are always larger than the PBE results
by 0.4--0.7 eV,
\red{consistent with the PBE+vdW and EX+cRPA results.\cite{PhysRevB.85.041403}}

\begin{table*}[t]
\caption{\label{tbl:ads_energy}
Adsorption energy of benzene on the Si(100) surface obtained with several vdW-DFs.
The results calculated self-consistently with the structure relaxted for each functional are
shown in panel ``SC-relaxed,''
while those calculated self-consistently (non-self-consistently) with the structure
fixed to the PBE geometries are shown in panel ``SC-fixed (non-SC).''
The adsorption energy of the more stable structure is highlighted with bold faces.
$\Delta E_{\rm ads}\equiv E_{\rm ads}{\rm (BF)}-E_{\rm ads}{\rm (TB)}$ denotes
the energy difference between the two adsorption structures.
All energies are in eV.
}
{
 \renewcommand\arraystretch{1.25}
\begin{tabular}[t]{llccccccc}
\hline
\hline
 & &PBE&vdW-DF1&vdW-DF2&optPBE-vdW&optB88-vdW&optB86b-vdW&rev-vdW-DF2\\
\hline
SC-relaxed&$E_{\rm ads}$(BF)&\phantom{+}0.75&\phantom{+}{\bf 0.90}&\phantom{+}{\bf 0.74}&\phantom{+}1.26&
\phantom{+}1.37&\phantom{+}1.46&\phantom{+}1.33\\
        &$E_{\rm ads}$(TB)&\phantom{+}{\bf 0.97}&\phantom{+}0.89&\phantom{+}0.60&\phantom{+}{\bf 1.36}&
\phantom{+}{\bf 1.53}&\phantom{+}{\bf 1.69}&\phantom{+}{\bf 1.55}\\
&$\Delta E_{\rm ads}$&$-0.23$&$+0.01$&$+0.15$&$-0.11$&$-0.16$&$-0.23$&$-0.23$\\
\hline
SC-fixed&$E_{\rm ads}$(BF)&&\phantom{+}0.90&\phantom{+}{\bf 0.74}&\phantom{+}1.25&
\phantom{+}1.37&\phantom{+}1.46&\phantom{+}1.32\\
&$E_{\rm ads}$(TB)&&\phantom{+}{\bf 0.91}&\phantom{+}0.63&\phantom{+}{\bf 1.37}&
\phantom{+}{\bf 1.54}&\phantom{+}{\bf 1.70}&\phantom{+}{\bf 1.56}\\
&$\Delta E_{\rm ads}$&&$-0.01$&$+0.11$&$-0.12$&$-0.17$&$-0.23$&$-0.23$\\
\hline
non-SC&$E_{\rm ads}$(BF)&&\phantom{+}0.89&\phantom{+}{\bf 0.73}&\phantom{+}1.25&
\phantom{+}1.38&\phantom{+}1.47&\phantom{+}1.33\\
&$E_{\rm ads}$(TB)&&\phantom{+}0.89&\phantom{+}0.58&\phantom{+}{\bf 1.37}&
\phantom{+}{\bf 1.54}&\phantom{+}{\bf 1.70}&\phantom{+}{\bf 1.56}\\
&$\Delta E_{\rm ads}$&&$\phantom{+}0.00$&$+0.15$&$-0.11$&$-0.17$&$-0.23$&$-0.23$\\
\hline
\hline
\end{tabular}
}
\end{table*}

We now investigate the relative stability of BF and TB for each vdW-DF
using the energy difference
$\Delta E_{\rm ads}\equiv E_{\rm ads}{\rm (BF)}-E_{\rm ads}{\rm (TB)}$.
For vdW-DF1, BF is marginally more stable than TB by $\Delta E_{\rm ads}=$
7 (2) meV in the SC-relaxed (non-SC) result,
while less stable by 7 meV in the SC-fixed result.
Although the energy difference is rather small,
the stabilization of BF by vdW-DF1 is qualitatively in 
agreement with the previous vdW-DF1 results by Johnston {\it et al}.\cite{PhysRevB.77.121404}
Note that in the non-SC treatment used in 
Ref.~\citenum{PhysRevB.77.121404},
the deformation energies of benzene and the Si(100) surface are calculated within PBE
to avoid the so-called eggbox effect inherent to the real-space vdW-DF method.
This treatment may be justified as long as the deformation energies can be well approximated
by the PBE values within the margin of error sufficiently smaller than
$\Delta E_{\rm ads}$.
In the Fourier-space vdW-DF method based on the RPS algorithm,
on the other hand, the deformation energies can be calculated on the same footing
without suffering from the eggbox effect.
The results shown in Table~\ref{tbl:ads_energy} are thus obtained
for all of the SC-relaxed, SC-fixed and non-SC treatments.
If we adopt the PBE deformation energies instead of the vdW-DF1 ones
in the non-SC results, we obtain $E_{\rm ads}=0.81$ (0.82) eV for BF (TB)
and BF becomes less stable than TB by 13 meV.
This indicates that the small energy difference of the order of 10 meV
can be easily affected by the detail of the estimation procedure of $E_{\rm ads}$.
Unlike the competing behavior of BF and TB in the vdW-DF1 results,
vdW-DF2 clearly supports BF with $\Delta E_{\rm ads}=0.15$ eV.
From the comparison with the PBE results,
it is tempting to consider that the relative stability of BF
in the vdW-DF2 results just stems from the destabilization of TB.
However, a more detailed analysis reveals that the adsorption energies of BF and TB
are determined by the balance between several
energy contributions as will be discussed in Sec.~\ref{sec:interaction-deformation}.

In sharp contrast to vdW-DF1 and vdW-DF2, the other vdW-DFs all prefer TB to BF,
agreeing with the
\red{PBE+vdW,\cite{PhysRevB.85.041403,Note1}
EX+cRPA\cite{PhysRevB.85.041403}
and DFT-D\cite{SurfSci.621.152} studies with $\Delta E_{\rm ads}=-0.16,-0.11$ and $-0.31$ eV,
respectively}.
\red{Thus the discrepancy between the vdW-DF method and other vdW-corrected approaches is resolved
by using recently proposed vdW-DFs with high accuracy.}
The three opt-vdW-DFs, i.e. optPBE-vdW, optB88-vdW
\cite{JPhysCondMatt.22.022201}
and optB86b-vdW,
\cite{PhysRevB.83.195131}
are obtained by replacing
the revPBE exchange in vdW-DF1 with optimized PBE, Becke88 (B88)
\cite{PhysRevA.38.3098}
and B86b exchanges.
Although wave functions and geometries are modified by the replacement of exchange
in the SC calculation,
the good agreement among the adsorption energies obtained with the SC-relaxed,
SC-fixed and non-SC method\red{s} strongly suggest
that these modifications
in fact make little difference in $E_{\rm ads}$.
From this observation, one may consider that the differences between the vdW-DF1
and opt-vdW-DFs results essentially derive from the difference in exchange energies.
Our results show that
optPBE-vdW, optB88-vdW and optB86b-vdW stabilizes TB more than BF by 0.1--0.2 eV
compared with vdW-DF1,
giving $\Delta E_{\rm ads}=-0.11, -0.17$ and $-0.23$ eV, respectively.
In the same way, the difference between the vdW-DF2 and rev-vdW-DF2 results
can also be considered as a result of the difference between
the PW86R and B86R exchanges used in these vdW-DFs.
rev-vdW-DF2 stabilizes TB more than BF by 0.36 eV compared with vdW-DF2,
which results in $\Delta E_{\rm ads}=-0.23$ eV.
Thus the relative stability of BF seen in vdW-DF1 and vdW-DF2 results is inverted
by the replacement of exchange with a less steeply rising $F_{\rm x}$.

It should be noted that opbB86b-vdW and rev-vdW-DF2 give similar adsorption energy
difference $\Delta E_{\rm ads}=-0.23$ eV.
Naively, the analogy between the two vdW-DFs can be expected from the fact
that optB86b and B86R exchanges derive from the same root.
\cite{JChemPhys.85.7184}
In addition, both optB86b and B86R exchange functionals match the gradient expansion
approximation in the slowing varying density limit.
Although $E_{\rm ads}$'s obtained with optB86b-vdW and rev-vdW-DF2 differ
because of different NL correlation used,
the results suggest that the exchange energy in the slowly varying density region plays an important
role in determining the stable structure of benzene on Si(100).

\begin{figure*}[t]
\includegraphics[width=.9\linewidth]{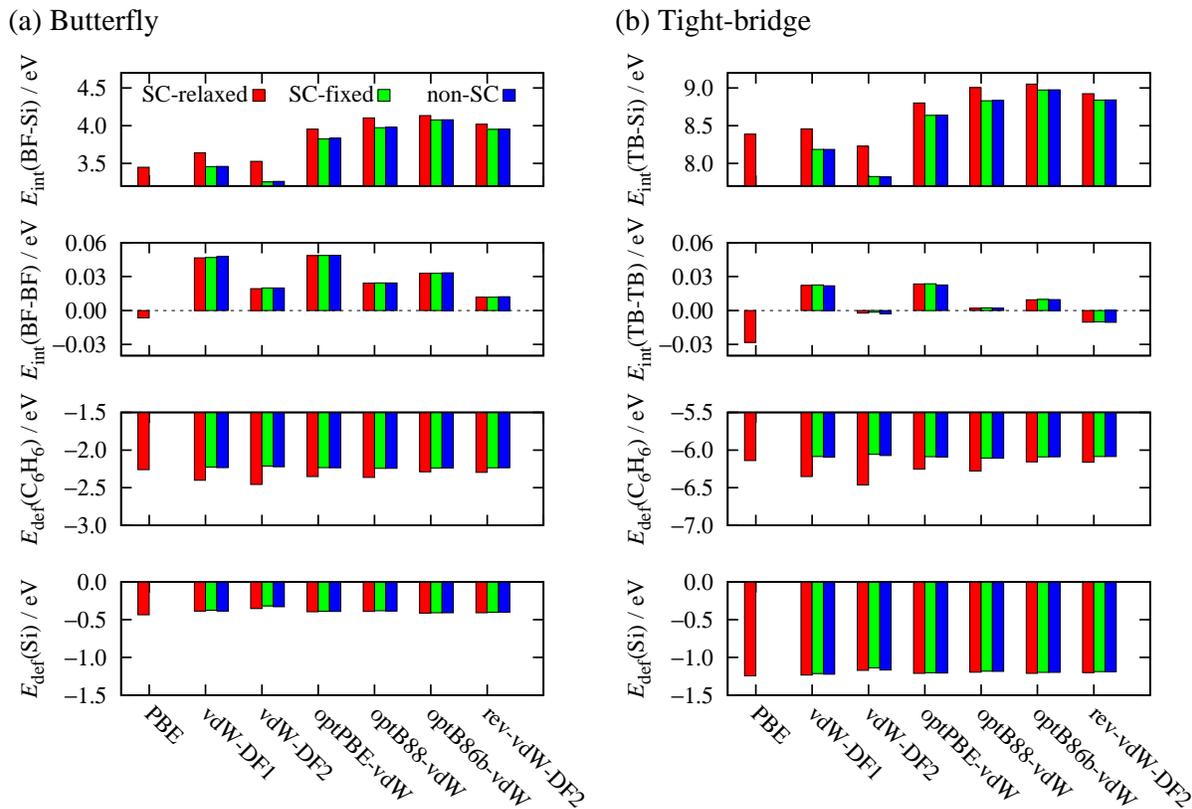}
\caption{\label{fig:energy_dec}(Color online)
Adsorption energy decomposed into four contributions:
benzene-surface interaction energy $E_{\rm int}$(BF/TB-Si),
benzene-benzene interaction energy $E_{\rm int}$(BF-BF/TB-TB),
benzene deformation energy $E_{\rm def}({\rm C_6H_6})$ and surface deformation energy $E_{\rm def}$(Si).
Panel (a) [(b)] shows the results for the BF (TB) structure.
The results calculated self-consistently with the structure relaxed for each functional are
shown in red, while those calculated self-consistently (non-self-consistently) with the structure
fixed to the PBE geometries are shown in green (blue).
Note that the energy scale of the benzene-benzene interaction is smaller than the others.}
\end{figure*}

\begin{table*}[t]
\caption{\label{tbl:bond}(Color online)
C--C, Si--Si and C--Si bond lengths in {\AA} of the BF and TB structures relaxed for each vdW-DF.
In the results of C--C (Si--Si) bond lengths, the values in brackets denote
the deviations from the results of an isolated benzene molecule
[a clean Si(100)-c($4\times2$) surface].
Redundant results for equivalent bonds are omitted.
The indices of C and Si atoms are shown in the schematics below,
where only the two highest Si layers are shown.
}
{
\renewcommand\arraystretch{1.25}
\begin{tabular}{llccccccc}
\hline
\hline
&&PBE&vdW-DF1&vdW-DF2&optPBE-vdW&optB88-vdW&optB86b-vdW&rev-vdW-DF2\\
\hline
${\rm C_6H_6}$&$l_{\rm C-C}$&1.396 \phantom{(+11.2\%)}&1.396 \phantom{(+11.2\%)}&
1.398 \phantom{(+11.2\%)}&1.396 \phantom{(+11.2\%)}&1.394 \phantom{(+11.2\%)}&
1.394 \phantom{(+11.2\%)}&1.396 \phantom{(+11.2\%)}\\
\hline
Si(100)&$l_{\rm Si_1-Si_2}$&2.349 \phantom{(+11.2\%)}&2.345 \phantom{(+11.2\%)}&2.346 \phantom{(+11.2\%)}&
2.346 \phantom{(+11.2\%)}&2.345 \phantom{(+11.2\%)}&2.349 \phantom{(+11.2\%)}&2.349 \phantom{(+11.2\%)}\\
&$l_{\rm Si_3-Si_4}$&2.405 \phantom{(+11.2\%)}&2.413 \phantom{(+11.2\%)}&2.415 \phantom{(+11.2\%)}&
2.413 \phantom{(+11.2\%)}&2.413 \phantom{(+11.2\%)}&2.412 \phantom{(+11.2\%)}&2.410 \phantom{(+11.2\%)}\\
&$l_{\rm Si_5-Si_6}$&2.363 \phantom{(+11.2\%)}&2.371 \phantom{(+11.2\%)}&2.378 \phantom{(+11.2\%)}&
2.374 \phantom{(+11.2\%)}&2.375 \phantom{(+11.2\%)}&2.372 \phantom{(+11.2\%)}&2.370 \phantom{(+11.2\%)}\\
\hline
BF&
$l_{\rm C_1-C_2}$&1.500 (+\phantom{1}7.5\%)&1.509 (+\phantom{1}8.1\%)&1.515 (+\phantom{1}8.4\%)&
1.505 (+\phantom{1}7.8\%)&1.504 (+\phantom{1}7.9\%)&1.500 (+\phantom{1}7.6\%)&1.501 (+\phantom{1}7.6\%)\\
&
$l_{\rm C_3-C_4}$&1.346 ($-$\phantom{1}3.6\%)&1.344 ($-$\phantom{1}3.7\%)&1.346 ($-$\phantom{1}3.8\%)&
1.345 ($-$\phantom{1}3.6\%)&1.344 ($-$\phantom{1}3.6\%)&1.346 ($-$\phantom{1}3.4\%)&1.347 ($-$\phantom{1}3.5\%)\\
&
$l_{\rm C_5-C_6}$&1.498 (+\phantom{1}6.9\%)&1.507 (+\phantom{1}7.4\%)&1.513 (+\phantom{1}7.6\%)&
1.503  (+\phantom{1}7.2\%)&1.503  (+\phantom{1}7.2\%)&1.499  (+\phantom{1}7.0\%)&1.500 (+\phantom{1}6.9\%)\\
\cline{2-9}
&$l_{\rm C_2-Si_3}$&1.971 \phantom{(+11.2\%)}&1.971 \phantom{(+11.2\%)}&1.975 \phantom{(+11.2\%)}&
1.971 \phantom{(+11.2\%)}&1.968 \phantom{(+11.2\%)}&1.969 \phantom{(+11.2\%)}&1.970 \phantom{(+11.2\%)}\\
&$l_{\rm C_5-Si_8}$&1.991 \phantom{(+11.2\%)}&1.992 \phantom{(+11.2\%)}&1.999 \phantom{(+11.2\%)}&
1.993 \phantom{(+11.2\%)}&1.990 \phantom{(+11.2\%)}&1.990 \phantom{(+11.2\%)}&1.991 \phantom{(+11.2\%)}\\
\cline{2-9}
&$l_{\rm Si_1-Si_2}$&2.346 ($-$\phantom{1}0.2\%)&2.346 (+\phantom{1}0.1\%)&2.348 (+\phantom{1}0.1\%)&
2.349 (+\phantom{1}0.1\%)&2.349 (+\phantom{1}0.2\%)&2.350 (+\phantom{1}0.0\%)&2.350 (+\phantom{1}0.0\%)\\
&$l_{\rm Si_3-Si_4}$&2.364 ($-$\phantom{1}1.7\%)&2.360 ($-$\phantom{1}2.2\%)&2.360 ($-$\phantom{1}2.3\%)&
2.362 ($-$\phantom{1}2.1\%)&2.362 ($-$\phantom{1}2.1\%)&2.363 ($-$\phantom{1}2.0\%)&2.363 ($-$\phantom{1}2.0\%)\\
&$l_{\rm Si_5-Si_6}$&2.344 ($-$\phantom{1}0.8\%)&2.355 ($-$\phantom{1}0.7\%)&2.367 ($-$\phantom{1}0.5\%)&
2.354 ($-$\phantom{1}0.8\%)&2.353 ($-$\phantom{1}0.9\%)&2.352 ($-$\phantom{1}0.8\%)&2.353 ($-$\phantom{1}0.8\%)\\
&$l_{\rm Si_7-Si_8}$&2.378 (+\phantom{1}1.2\%)&2.373 (+\phantom{1}1.2\%)&2.372 (+\phantom{1}1.1\%)&
2.376 (+\phantom{1}1.3\%)&2.376 (+\phantom{1}1.3\%)&2.377 (+\phantom{1}1.2\%)&2.377 (+\phantom{1}1.2\%)\\
&$l_{\rm Si_9-Si_{10}}$&2.398 ($-$\phantom{1}0.3\%)&2.397 ($-$\phantom{1}0.6\%)&2.402 ($-$\phantom{1}0.6\%)&
2.399 ($-$\phantom{1}0.5\%)&2.400 ($-$\phantom{1}0.5\%)&2.401 ($-$\phantom{1}0.5\%)&2.401 ($-$\phantom{1}0.4\%)\\
&$l_{\rm Si_3-Si_8}$&2.391 (+\phantom{1}1.2\%)&2.396 (+\phantom{1}1.0\%)&2.408 (+\phantom{1}1.3\%)&
2.394 (+\phantom{1}0.9\%)&2.394 (+\phantom{1}0.8\%)&2.391 (+\phantom{1}0.8\%)&2.392 (+\phantom{1}0.9\%)\\
\hline
TB&
$l_{\rm C_1-C_2}$&1.572 (+12.7\%)&1.587 (+13.7\%)&1.597 (+14.2\%)&
1.581 (+13.3\%)&1.580 (+13.3\%)&1.574 (+12.9\%)&1.574 (+12.8\%)\\
&$l_{\rm C_2-C_3}$&1.497 (+\phantom{1}7.3\%)&1.505 (+\phantom{1}7.8\%)&1.508 (+\phantom{1}7.9\%)&
1.501 (+\phantom{1}7.5\%)&1.500 (+\phantom{1}7.6\%)&1.497 (+\phantom{1}7.4\%)&1.497 (+\phantom{1}7.3\%)\\
&$l_{\rm C_3-C_4}$&1.347 ($-$\phantom{1}3.5\%)&1.347 ($-$\phantom{1}3.5\%)&1.348 ($-$\phantom{1}3.6\%)&
1.347 ($-$\phantom{1}3.5\%)&1.346 ($-$\phantom{1}3.4\%)&1.348 ($-$\phantom{1}3.4\%)&1.348 ($-$\phantom{1}3.4\%)\\
&
$l_{\rm C_6-C_1}$&1.575 (+11.4\%)&1.590 (+12.2\%)&1.600 (+12.6\%)&1.583 (+11.8\%)&1.582 (+11.9\%)&
1.577 (+11.6\%)&1.578 (+11.5\%)\\
\cline{2-9}
&$l_{\rm C_1-Si_1}$&2.017 \phantom{(+11.2\%)}&2.014 \phantom{(+11.2\%)}&2.016 \phantom{(+11.2\%)}&
2.017 \phantom{(+11.2\%)}&2.015 \phantom{(+11.2\%)}&2.017 \phantom{(+11.2\%)}&2.017 \phantom{(+11.2\%)}\\
&$l_{\rm C_2-Si_3}$&1.993 \phantom{(+11.2\%)}&1.989 \phantom{(+11.2\%)}&1.991 \phantom{(+11.2\%)}&
1.990 \phantom{(+11.2\%)}&1.987 \phantom{(+11.2\%)}&1.989 \phantom{(+11.2\%)}&1.989 \phantom{(+11.2\%)}\\
\cline{2-9}
&$l_{\rm Si_1-Si_2}$&2.338 ($-$\phantom{1}0.5\%)&2.337 ($-$\phantom{1}0.3\%)&2.341 ($-$\phantom{1}0.2\%)&
2.339 ($-$\phantom{1}0.3\%)&2.339 ($-$\phantom{1}0.2\%)&2.340 ($-$\phantom{1}0.4\%)&2.340 ($-$\phantom{1}0.4\%)\\
&$l_{\rm Si_2-Si_3}$&2.341 ($-$\phantom{1}2.6\%)&2.338 ($-$\phantom{1}3.1\%)&2.339 ($-$\phantom{1}3.1\%)&
2.340 ($-$\phantom{1}3.0\%)&2.340 ($-$\phantom{1}3.0\%)&2.341 ($-$\phantom{1}2.9\%)&2.341 ($-$\phantom{1}2.9\%)\\
&$l_{\rm Si_3-Si_4}$&2.427 (+\phantom{1}0.9\%)&2.427 (+\phantom{1}0.6\%)&2.426 (+\phantom{1}0.4\%)&
2.429 (+\phantom{1}0.7\%)&2.429 (+\phantom{1}0.7\%)&2.429 (+\phantom{1}0.7\%)&2.429 (+\phantom{1}0.8\%)\\
&$l_{\rm Si_4-Si_5}$&2.438 (+\phantom{1}3.8\%)&2.439 (+\phantom{1}4.0\%)&2.438 (+\phantom{1}3.9\%)&
2.442 (+\phantom{1}4.1\%)&2.442 (+\phantom{1}4.1\%)&2.441 (+\phantom{1}3.9\%)&2.442 (+\phantom{1}3.9\%)\\
&$l_{\rm Si_5-Si_6}$&2.352 ($-$\phantom{1}0.4\%)&2.358 ($-$\phantom{1}0.5\%)&2.367 ($-$\phantom{1}0.5\%)&
2.358 ($-$\phantom{1}0.6\%)&2.359 ($-$\phantom{1}0.7\%)&2.358 ($-$\phantom{1}0.6\%)&2.359 ($-$\phantom{1}0.5\%)\\
&$l_{\rm Si_3-Si_8}$&2.384 (+\phantom{1}0.9\%)&2.392 (+\phantom{1}0.9\%)&2.403 (+\phantom{1}1.0\%)&
2.389 (+\phantom{1}0.7\%)&2.389 (+\phantom{1}0.6\%)&2.386 (+\phantom{1}0.6\%)&2.387 (+\phantom{1}0.7\%)\\
\hline
\hline
 & & \\
\end{tabular}
}\\
\hfill\includegraphics[width=.25\linewidth]{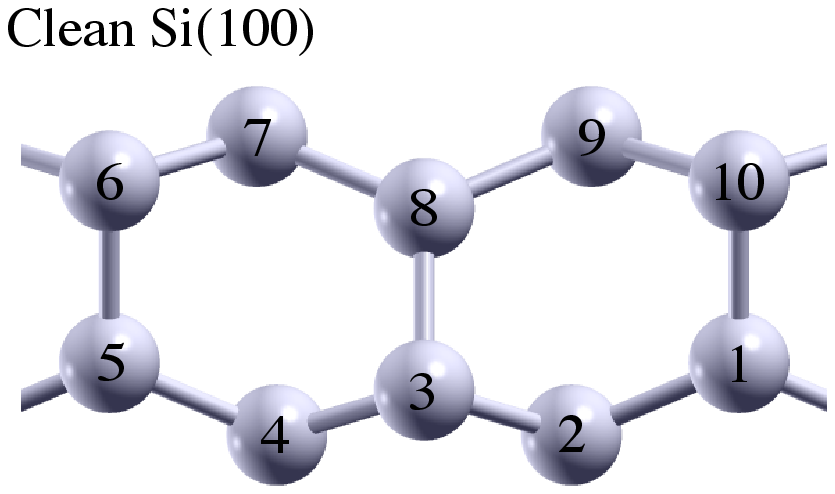}\hfill
\includegraphics[width=.25\linewidth]{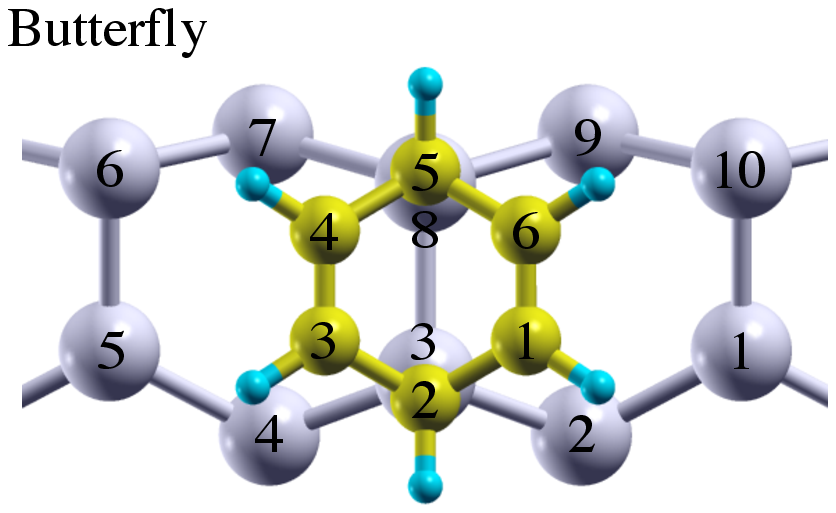}\hfill
\includegraphics[width=.25\linewidth]{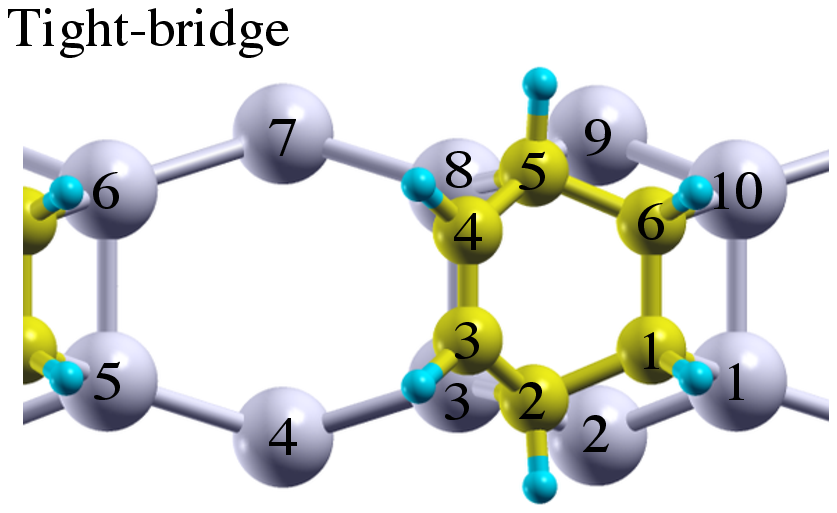}\hfill{}
\end{table*}

\subsection{Energy decomposition analysis}
\label{sec:interaction-deformation}
To identify the origin of the similarities
and differences among the results obtained with different vdW-DFs,
we divide $E_{\rm ads}$ into interaction
and deformation energies as shown in Fig.~\ref{fig:energy_dec},
following the procedure used in 
Refs.~\citenum{PhysRevB.77.121404} and \citenum{PhysRevB.85.041403}:
\begin{align}
 E_{\rm ads}&=E_{\rm int}(\mbox{BF/TB-Si}) + E_{\rm int}(\mbox{\rm BF-BF/TB-TB})\nonumber\\
&\phantom{=}+E_{\rm def}({\rm C}_{6}{\rm H}_{6})+E_{\rm def}({\rm Si}).
\end{align}
Here $E_{\rm int}$(BF-Si) [$E_{\rm int}$(TB-Si)] denotes the interaction energy
between the benzene layer in the BF (TB) configuration and the Si(100) surface.
This is calculated from the energy difference between the adsorbed system
and the reference systems,
in which the geometries of benzene molecules and the Si(100) surface
are fixed to the adsorption structures.
$E_{\rm int}$(BF-BF) [$E_{\rm int}$(TB-TB)] is the interaction energy
between benzene molecules in the BF (TB) configuration, and is calculated similarly
using the geometries fixed to the adsorption structures.
The deformation energy $E_{\rm def}({\rm C_6H_6})$ [$E_{\rm def}({\rm Si})$] is defined
as the energy loss to deform the benzene molecule [Si(100) surface] upon adsorption.

From the comparison among the SC-relaxed, SC-fixed and non-SC results of
the four energy contributions,
one can see that the SC-fixed and non-SC results are almost the same,
whereas the SC-relaxed results show small but notable deviations from the others
especially in the leading contributions, $E_{\rm int}$(BF/TB-Si)
and $E_{\rm def}({\rm C_6H_6})$.
This reveals that the SC calculation does not significantly modify these four energies
as long as the geometries are fixed, validating the previous non-SC vdW-DF calculations.
On the other hand, the deviations seen in the SC-relaxed results are due largely
to structure relaxation, hence reflecting the characteristics of each vdW-DF.
For example, the optB86b-vdW and rev-vdW-DF2 results of $E_{\rm def}({\rm C_6H_6})$
are analogous to the PBE ones both for BF and TB,
although their SC-relaxed results show slight deviations $\lesssim 0.1$ eV
from the SC-fixed and non-SC ones.
The analogy to the PBE results is expected from the fact that $E_{\rm def}({\rm C_6H_6})$
essentially derives from intra-molecular interaction,
and also holds for the SC-fixed and non-SC results of the other vdW-DFs.
However, the SC-relaxed results of the other vdW-DFs display larger deviations
from the unrelaxed ones by 0.1-0.4 eV,
which is more pronounced for TB than BF.

To gain more insight into the functional dependence seen in $E_{\rm def}({\rm C_6H_6})$,
we show in Table~\ref{tbl:bond} the C--C bond lengths of benzene
for the two adsorption structures relaxed with each vdW-DF.
One can see that the C--C bonds are significantly modified upon adsorption
as a result of the distortion of benzene in the respective adsorption structures.
The shortened bonds little depend on the functional both for BF and TB
and slightly longer than the typical length of a C=C double bond $\simeq 1.33$ \AA,
suggesting that double bonds are nearly formed between the C atoms not bonded
to the Si atoms on the surface.
On the other hand, the elongated bonds exhibit a clearer functional dependence
analogous to $E_{\rm def}({\rm C_6H_6})$ mentioned above.
That is, the C--C bond lengths obtained with optB86b-vdW and rev-vdW-DF2 are similar
to the PBE results, while the other vdW-DFs give longer C--C bonds.
In particular, vdW-DF1 and vdW-DF2 give largest increases in the C--C bond lengths,
which is consistent with the repulsive nature of the exchange functionals used in these vdW-DFs.
From the analogy between the behaviors of $E_{\rm def}({\rm C_6H_6})$ and C--C bond lengths,
the functional dependence seen in the SC-relaxed results of $E_{\rm def}({\rm C_6H_6})$
can be attributed to the extension of the C--C bonds as a result of structure relaxation.

The benzene-surface interaction energy for TB, $E_{\rm int}$(TB-Si),
is more than twice larger than that for BF,
compensating the larger energy loss from $E_{\rm def}({\rm C_6H_6})$ for TB.
The larger interaction can be understood from the fact the benzene molecule
interacts with two Si dimers (a single Si dimer) in the TB (BF) configuration.
More importantly, $E_{\rm int}$(BF/TB-Si) shows a significant functional
dependence as compared with $E_{\rm def}({\rm C}_6{\rm H}_6)$,
since NL correlation plays an important role in the interaction.
The results of the three opt-vdW-DFs and rev-vdW-DF2 consistently display increases
from the PBE ones
by 0.3-0.7 eV.
On the other hand, the SC-fixed and non-SC results obtained with vdW-DF1 and vdW-DF2 are
similar to or smaller than the PBE results
despite the presence of attractive NL correlation.
Although the $E_{\rm int}$(BF/TB-Si) is increased by 0.2-0.3 eV in the SC-relaxed results,
they are still closer to the PBE ones than to those of the other vdW-DFs.
The smaller benzene-surface interaction seen in the vdW-DF1 and vdW-DF2 results
cannot be ascribed to the C--Si bonds between benzene and the Si surface,
since the lengths of these bonds show less functional dependence than the C--C bonds
discussed above.
Rather, the attraction due to the NL correlation is counteracted by the too repulsive exchanges
used in these vdW-DFs.
This can be understood in particular from the comparison of the non-SC results,
since the energy difference between vdW-DF1 and three opt-vdW-DFs or
between vdW-DF2 and rev-vdW-DF2 genuinely stems from the difference in the exchange part.

In contrast to the previous two energy contributions,
the surface deformation energy $E_{\rm def}$(Si) shows little difference among the SC-relaxed,
SC-fixed and non-SC results, suggesting that structure relaxation plays only a minor role
in the deformation of the Si(100) surface.
This can be confirmed by examining the Si--Si bond lengths near the Si(100) surface
shown in Table~\ref{tbl:bond}.
One can see that the Si--Si bonds reflect the deformation
of the surface Si dimers due to benzene adsorption
and show slight dependence on functional.
However, the change in the Si--Si bond lengths
is much smaller than that in the C--C bond lengths.
As a result $E_{\rm def}$(Si) shows the relatively monotonic behavior
with little dependence on structure relaxation and functional.
A characteristic feature of $E_{\rm def}$(Si) is that the energy loss for TB
is roughly three times larger than that for BF,
which results from the fact that the Si--Si bonds for TB is more extended
than those for BF.

The benzene-benzene interaction energy $E_{\rm int}$(BF-BF/TB-TB)
is even less affected by structure relaxation,
suggesting that the detailed structure of benzene has little
influence on the inter-molecular interaction.
One can see that the interaction acts repulsively within PBE,
while it becomes attractive or less repulsive for vdW-DFs.
In particular, one finds that BF is always more attractive than TB by 0.02--0.03 eV,
which is qualitatively consistent with the experiments.
\cite{PhysRevB.71.115311, JPhysCondMatt.20.304206}
Note, however, that $E_{\rm int}$(BF-BF/TB-TB) is smaller than the other contributions
by a few orders of magnitude.
This strongly suggests that the inter-molecular interaction cannot be a main origin
of the relative stability of BF even in the presence of NL correlation,
since similar energy difference has been already obtained within PBE.

We stress that despite structure relaxation has a significant influence
on $E_{\rm int}$(BF-Si/TB-Si) and $E_{\rm def}({\rm C_6H_6})$,
$E_{\rm ads}$ shows only a slight difference
among the SC-relaxed, SC-fixed and non-SC results.
This means that the errors in the latter two treatments are canceled out,
\cite{JChemPhys.128.244704}
which results in the good agreement with the
\red{SC-relaxed} results of $E_{\rm ads}$.
Thus, in order to guarantee full cancellation of such errors,
the four energy contributions should be calculated on the same footing.
In the non-SC vdW-DF method of Ref.~\citenum{PhysRevB.77.121404}, 
however, the deformation energies are estimated by PBE instead of vdW-DF1,
which could affect the error cancellation.
Indeed, our non-SC results show that the differences between the deformation energies
obtained
\red{with} PBE and vdW-DF1 are apparently rather small but still large enough
to influence the subtle difference between BF and TB in the vdW-DF1 results.

Finally, we interpret the functional dependence of $E_{\rm ads}$
in terms of the four energy contributions, focusing on the SC-relaxed results.
Our results demonstrate that $E_{\rm ads}$ is essentially characterized
by the balance between the two leading contributions,
$E_{\rm int}$(BF/TB-Si) and $E_{\rm def}({\rm C_6H_6})$.
In the case of opt-vdW-DFs and rev-vdW-DF2, $E_{\rm def}({\rm C_6H_6})$'s are
similar to the PBE ones. 
In addition, the benzene-surface interaction becomes larger than the PBE one
due to the less repulsive exchange and the attractive NL correlation,
leading to the increase in $E_{\rm ads}$ both for BF and TB.
As a result, relative stability is unchanged from the PBE case
for these vdW-DFs,
hence the TB remains more stable than BF.
For vdW-DF1 and vdW-DF2, on the other hand,
TB is more destabilized than BF through $E_{\rm def}({\rm C_6H_6})$,
and in addition $E_{\rm int}$(BF-Si) increases more than $E_{\rm int}$(TB-Si)
compared with the PBE result.
Thus both the two leading contributions act to stabilize BF compared with TB,
which is the origin of the relative stability of BF found in the vdW-DF1 and vdW-DF2 results.

\section{Conclusion}
We have theoretically investigated the adsorption structure of benzene
on the Si(100) surface at saturation coverage
using the SC vdW-DF method based on several vdW-DFs.
Our results show that 
\red{recently proposed vdW-DFs with high accuracy all predict that TB is more stable than BF
in good agreement with EX+cRPA and other vdW-corrected calculations,
in sharp contrast to the vdW-DF1 (vdW-DF2) results that marginally (robustly) prefer BF to TB.
}
The relative stability between BF and TB has been analyzed in terms of interaction
and deformation energies of benzene and the Si(100) surface.
The functional dependence of the relative stability of BF and TB
\red{is determined by} the balance 
\red{between the} two leading contributions,
benzene-surface interaction and benzene deformation energies,
\red{both of which act to destabilize TB as compared with BF for vdW-DF1 and vdW-DF2,
because they underestimate the strength of the covalent bonding severely.
\cite{PhysRevB.83.195131}}
\red{Thus we conclude that TB is the most stable adsorption structure of benzene on Si(001)
at saturation coverage and zero temperature.}
\red{Further theoretical investigation is required to resolve the controversy in the present system},
since a variety of experiments still indicate the relative stability of BF
especially at quasi-saturation coverage.
Consideration of steric hinderance
\cite{PhysRevB.72.235317, SurfSci.621.152}
and thermal vibration
\cite{PhysRevB.85.041403}
can be promising approaches beyond simple energetics at zero temperature.
Nevertheless, the present study has demonstrated that benzene on Si(100)
can be a benchmark system to assess the performance of new functionals
in the sense that it is of critical importance to describe both the covalent and vdW bonding
very accurately in order to predict the relative stability of the adsorption structures.

\acknowledgments
Y.~H. thanks Nicolae Atodiresei and Vasile Caciuc for valuable discussions.
I.~H. acknowledges financial support from Ministry of Education, Culture, Sports,
Science and Technology in Japan (MEXT) through World Premier International Research Center
Initiative for Materials Nanoarchitectonics (WPI-MANA) and ``Development of Environmental 
Technology using Nanotechnology'' program.
This work has been partly supported by Grant-in-Aid for Young Scientists (B) (No.~15K17682)
from Japan Society for the Promotion of Science (JSPS),
Grant-in-Aid for Scientific Research on Innovative Areas ``Molecular Architectnics:
Orchestration of Single Molecules for Novel Functions'' (No.~25110006) from MEXT,
Grant-in-Aid for Scientific Research on Innovative Areas ``3D Active-Site Science''
(No. 26105010 and No. 26105011) from JSPS,
the JST ACT-C program, the MEXT ``Elements  Strategy Initiative to Form Core Research Center'' program,
and the JSPS Core-to-Core Program (Type A) ``Advanced Research Networks:
Computational Materials Design on Green Energy.''
The computation in this work has been done with the facilities of Supercomputer Center,
Institute for Solid State Physics, University of Tokyo.


\begin{thebibliography}{100}%
\makeatletter
\providecommand \@ifxundefined [1]{%
 \@ifx{#1\undefined}
}%
\providecommand \@ifnum [1]{%
 \ifnum #1\expandafter \@firstoftwo
 \else \expandafter \@secondoftwo
 \fi
}%
\providecommand \@ifx [1]{%
 \ifx #1\expandafter \@firstoftwo
 \else \expandafter \@secondoftwo
 \fi
}%
\providecommand \natexlab [1]{#1}%
\providecommand \enquote  [1]{``#1''}%
\providecommand \bibnamefont  [1]{#1}%
\providecommand \bibfnamefont [1]{#1}%
\providecommand \citenamefont [1]{#1}%
\providecommand \href@noop [0]{\@secondoftwo}%
\providecommand \href [0]{\begingroup \@sanitize@url \@href}%
\providecommand \@href[1]{\@@startlink{#1}\@@href}%
\providecommand \@@href[1]{\endgroup#1\@@endlink}%
\providecommand \@sanitize@url [0]{\catcode `\\12\catcode `\$12\catcode
  `\&12\catcode `\#12\catcode `\^12\catcode `\_12\catcode `\%12\relax}%
\providecommand \@@startlink[1]{}%
\providecommand \@@endlink[0]{}%
\providecommand \url  [0]{\begingroup\@sanitize@url \@url }%
\providecommand \@url [1]{\endgroup\@href {#1}{\urlprefix }}%
\providecommand \urlprefix  [0]{URL }%
\providecommand \Eprint [0]{\href }%
\providecommand \doibase [0]{http://dx.doi.org/}%
\providecommand \selectlanguage [0]{\@gobble}%
\providecommand \bibinfo  [0]{\@secondoftwo}%
\providecommand \bibfield  [0]{\@secondoftwo}%
\providecommand \translation [1]{[#1]}%
\providecommand \BibitemOpen [0]{}%
\providecommand \bibitemStop [0]{}%
\providecommand \bibitemNoStop [0]{.\EOS\space}%
\providecommand \EOS [0]{\spacefactor3000\relax}%
\providecommand \BibitemShut  [1]{\csname bibitem#1\endcsname}%
\let\auto@bib@innerbib\@empty
\bibitem [{\citenamefont {{Taguchi}}\ \emph {et~al.}(1991)\citenamefont
  {{Taguchi}}, \citenamefont {{Fujisawa}}, \citenamefont {{Takaoka}},
  \citenamefont {{Okada}},\ and\ \citenamefont {{Nishijima}}}]{JChemPhys.95.1}%
  \BibitemOpen
  \bibfield  {author} {\bibinfo {author} {\bibfnamefont {Y.}~\bibnamefont
  {{Taguchi}}}, \bibinfo {author} {\bibfnamefont {M.}~\bibnamefont
  {{Fujisawa}}}, \bibinfo {author} {\bibfnamefont {T.}~\bibnamefont
  {{Takaoka}}}, \bibinfo {author} {\bibfnamefont {T.}~\bibnamefont {{Okada}}},
  \ and\ \bibinfo {author} {\bibfnamefont {M.}~\bibnamefont {{Nishijima}}},\
  }\href {\doibase 10.1063/1.461498} {\bibfield  {journal} {\bibinfo  {journal}
  {J. Chem. Phys.}\ }\textbf {\bibinfo {volume} {95}},\ \bibinfo {pages} {6870}
  (\bibinfo {year} {1991})}\BibitemShut {NoStop}%
\bibitem [{\citenamefont {{Lopinski}}\ \emph
  {et~al.}(1998{\natexlab{a}})\citenamefont {{Lopinski}}, \citenamefont
  {{Moffatt}},\ and\ \citenamefont {{Wolkow}}}]{ChemPhysLett.282.305}%
  \BibitemOpen
  \bibfield  {author} {\bibinfo {author} {\bibfnamefont {G.~P.}\ \bibnamefont
  {{Lopinski}}}, \bibinfo {author} {\bibfnamefont {D.~J.}\ \bibnamefont
  {{Moffatt}}}, \ and\ \bibinfo {author} {\bibfnamefont {R.~A.}\ \bibnamefont
  {{Wolkow}}},\ }\href {\doibase 10.1016/S0009-2614(97)01317-1} {\bibfield
  {journal} {\bibinfo  {journal} {Chem. Phys. Lett.}\ }\textbf {\bibinfo
  {volume} {282}},\ \bibinfo {pages} {305} (\bibinfo {year}
  {1998}{\natexlab{a}})}\BibitemShut {NoStop}%
\bibitem [{\citenamefont {{Self}}\ \emph {et~al.}(1998)\citenamefont {{Self}},
  \citenamefont {{Pelzel}}, \citenamefont {{Owen}}, \citenamefont {{Yan}},
  \citenamefont {{Widdra}},\ and\ \citenamefont
  {{Weinberg}}}]{JVacSciTechnolA.16.1031}%
  \BibitemOpen
  \bibfield  {author} {\bibinfo {author} {\bibfnamefont {K.~W.}\ \bibnamefont
  {{Self}}}, \bibinfo {author} {\bibfnamefont {R.~I.}\ \bibnamefont
  {{Pelzel}}}, \bibinfo {author} {\bibfnamefont {J.~H.~G.}\ \bibnamefont
  {{Owen}}}, \bibinfo {author} {\bibfnamefont {C.}~\bibnamefont {{Yan}}},
  \bibinfo {author} {\bibfnamefont {W.}~\bibnamefont {{Widdra}}}, \ and\
  \bibinfo {author} {\bibfnamefont {W.~H.}\ \bibnamefont {{Weinberg}}},\
  }\href@noop {} {\bibfield  {journal} {\bibinfo  {journal} {J. Vac. Sci.
  Technol. A}\ }\textbf {\bibinfo {volume} {16}},\ \bibinfo {pages} {1031}
  (\bibinfo {year} {1998})}\BibitemShut {NoStop}%
\bibitem [{\citenamefont {{Lopinski}}\ \emph
  {et~al.}(1998{\natexlab{b}})\citenamefont {{Lopinski}}, \citenamefont
  {{Fortier}}, \citenamefont {{Moffatt}},\ and\ \citenamefont
  {{Wolkow}}}]{JVacSciTechnolA.16.1037}%
  \BibitemOpen
  \bibfield  {author} {\bibinfo {author} {\bibfnamefont {G.~P.}\ \bibnamefont
  {{Lopinski}}}, \bibinfo {author} {\bibfnamefont {T.~M.}\ \bibnamefont
  {{Fortier}}}, \bibinfo {author} {\bibfnamefont {D.~J.}\ \bibnamefont
  {{Moffatt}}}, \ and\ \bibinfo {author} {\bibfnamefont {R.~A.}\ \bibnamefont
  {{Wolkow}}},\ }\href@noop {} {\bibfield  {journal} {\bibinfo  {journal} {J.
  Vac. Sci. Technol. A}\ }\textbf {\bibinfo {volume} {16}},\ \bibinfo {pages}
  {1037} (\bibinfo {year} {1998}{\natexlab{b}})}\BibitemShut {NoStop}%
\bibitem [{\citenamefont {Borovsky}\ \emph {et~al.}(1998)\citenamefont
  {Borovsky}, \citenamefont {Krueger},\ and\ \citenamefont
  {Ganz}}]{PhysRevB.57.R4269}%
  \BibitemOpen
  \bibfield  {author} {\bibinfo {author} {\bibfnamefont {B.}~\bibnamefont
  {Borovsky}}, \bibinfo {author} {\bibfnamefont {M.}~\bibnamefont {Krueger}}, \
  and\ \bibinfo {author} {\bibfnamefont {E.}~\bibnamefont {Ganz}},\ }\href
  {\doibase 10.1103/PhysRevB.57.R4269} {\bibfield  {journal} {\bibinfo
  {journal} {Phys. Rev. B}\ }\textbf {\bibinfo {volume} {57}},\ \bibinfo
  {pages} {R4269} (\bibinfo {year} {1998})}\BibitemShut {NoStop}%
\bibitem [{\citenamefont {{Gokhale}}\ \emph {et~al.}(1998)\citenamefont
  {{Gokhale}}, \citenamefont {{Trischberger}}, \citenamefont {{Menzel}},
  \citenamefont {{Widdra}}, \citenamefont {{Dr{\"o}ge}}, \citenamefont
  {{Steinr{\"u}ck}}, \citenamefont {{Birkenheuer}}, \citenamefont
  {{Gutdeutsch}},\ and\ \citenamefont {{R{\"o}sch}}}]{JChemPhys.108.5554}%
  \BibitemOpen
  \bibfield  {author} {\bibinfo {author} {\bibfnamefont {S.}~\bibnamefont
  {{Gokhale}}}, \bibinfo {author} {\bibfnamefont {P.}~\bibnamefont
  {{Trischberger}}}, \bibinfo {author} {\bibfnamefont {D.}~\bibnamefont
  {{Menzel}}}, \bibinfo {author} {\bibfnamefont {W.}~\bibnamefont {{Widdra}}},
  \bibinfo {author} {\bibfnamefont {H.}~\bibnamefont {{Dr{\"o}ge}}}, \bibinfo
  {author} {\bibfnamefont {H.-P.}\ \bibnamefont {{Steinr{\"u}ck}}}, \bibinfo
  {author} {\bibfnamefont {U.}~\bibnamefont {{Birkenheuer}}}, \bibinfo {author}
  {\bibfnamefont {U.}~\bibnamefont {{Gutdeutsch}}}, \ and\ \bibinfo {author}
  {\bibfnamefont {N.}~\bibnamefont {{R{\"o}sch}}},\ }\href {\doibase
  10.1063/1.475945} {\bibfield  {journal} {\bibinfo  {journal} {J. Chem.
  Phys.}\ }\textbf {\bibinfo {volume} {108}},\ \bibinfo {pages} {5554}
  (\bibinfo {year} {1998})}\BibitemShut {NoStop}%
\bibitem [{\citenamefont {{Kong}}\ \emph {et~al.}(1998)\citenamefont {{Kong}},
  \citenamefont {{Teplyakov}}, \citenamefont {{Lyubovitsky}},\ and\
  \citenamefont {{Bent}}}]{SurfSci.411.286}%
  \BibitemOpen
  \bibfield  {author} {\bibinfo {author} {\bibfnamefont {M.~J.}\ \bibnamefont
  {{Kong}}}, \bibinfo {author} {\bibfnamefont {A.~V.}\ \bibnamefont
  {{Teplyakov}}}, \bibinfo {author} {\bibfnamefont {J.~G.}\ \bibnamefont
  {{Lyubovitsky}}}, \ and\ \bibinfo {author} {\bibfnamefont {S.~F.}\
  \bibnamefont {{Bent}}},\ }\href {\doibase 10.1016/S0039-6028(98)00336-7}
  {\bibfield  {journal} {\bibinfo  {journal} {Surf. Sci.}\ }\textbf {\bibinfo
  {volume} {411}},\ \bibinfo {pages} {286} (\bibinfo {year}
  {1998})}\BibitemShut {NoStop}%
\bibitem [{\citenamefont {{Li}}\ and\ \citenamefont
  {{Leung}}(2001)}]{SurfSci.479.69}%
  \BibitemOpen
  \bibfield  {author} {\bibinfo {author} {\bibfnamefont {Q.}~\bibnamefont
  {{Li}}}\ and\ \bibinfo {author} {\bibfnamefont {K.~T.}\ \bibnamefont
  {{Leung}}},\ }\href {\doibase 10.1016/S0039-6028(01)00958-X} {\bibfield
  {journal} {\bibinfo  {journal} {Surf. Sci.}\ }\textbf {\bibinfo {volume}
  {479}},\ \bibinfo {pages} {69} (\bibinfo {year} {2001})}\BibitemShut
  {NoStop}%
\bibitem [{\citenamefont {{Nagao}}\ \emph {et~al.}(2002)\citenamefont
  {{Nagao}}, \citenamefont {{Yamashita}}, \citenamefont {{Machida}},
  \citenamefont {{Hamaguchi}}, \citenamefont {{Yasui}}, \citenamefont
  {{Mukai}},\ and\ \citenamefont {{Yoshinobu}}}]{SurfSci.513.413}%
  \BibitemOpen
  \bibfield  {author} {\bibinfo {author} {\bibfnamefont {M.}~\bibnamefont
  {{Nagao}}}, \bibinfo {author} {\bibfnamefont {Y.}~\bibnamefont
  {{Yamashita}}}, \bibinfo {author} {\bibfnamefont {S.}~\bibnamefont
  {{Machida}}}, \bibinfo {author} {\bibfnamefont {K.}~\bibnamefont
  {{Hamaguchi}}}, \bibinfo {author} {\bibfnamefont {F.}~\bibnamefont
  {{Yasui}}}, \bibinfo {author} {\bibfnamefont {K.}~\bibnamefont {{Mukai}}}, \
  and\ \bibinfo {author} {\bibfnamefont {J.}~\bibnamefont {{Yoshinobu}}},\
  }\href {\doibase 10.1016/S0039-6028(02)01878-2} {\bibfield  {journal}
  {\bibinfo  {journal} {Surf. Sci.}\ }\textbf {\bibinfo {volume} {513}},\
  \bibinfo {pages} {413} (\bibinfo {year} {2002})}\BibitemShut {NoStop}%
\bibitem [{\citenamefont {Witkowski}\ \emph {et~al.}(2003)\citenamefont
  {Witkowski}, \citenamefont {Hennies}, \citenamefont {Pietzsch}, \citenamefont
  {Mattsson}, \citenamefont {F\"ohlisch}, \citenamefont {Wurth}, \citenamefont
  {Nagasono},\ and\ \citenamefont {Piancastelli}}]{PhysRevB.68.115408}%
  \BibitemOpen
  \bibfield  {author} {\bibinfo {author} {\bibfnamefont {N.}~\bibnamefont
  {Witkowski}}, \bibinfo {author} {\bibfnamefont {F.}~\bibnamefont {Hennies}},
  \bibinfo {author} {\bibfnamefont {A.}~\bibnamefont {Pietzsch}}, \bibinfo
  {author} {\bibfnamefont {S.}~\bibnamefont {Mattsson}}, \bibinfo {author}
  {\bibfnamefont {A.}~\bibnamefont {F\"ohlisch}}, \bibinfo {author}
  {\bibfnamefont {W.}~\bibnamefont {Wurth}}, \bibinfo {author} {\bibfnamefont
  {M.}~\bibnamefont {Nagasono}}, \ and\ \bibinfo {author} {\bibfnamefont
  {M.~N.}\ \bibnamefont {Piancastelli}},\ }\href {\doibase
  10.1103/PhysRevB.68.115408} {\bibfield  {journal} {\bibinfo  {journal} {Phys.
  Rev. B}\ }\textbf {\bibinfo {volume} {68}},\ \bibinfo {pages} {115408}
  (\bibinfo {year} {2003})}\BibitemShut {NoStop}%
\bibitem [{\citenamefont {{Naumkin}}\ \emph {et~al.}(2003)\citenamefont
  {{Naumkin}}, \citenamefont {{Polanyi}}, \citenamefont {{Rogers}},
  \citenamefont {{Hofer}},\ and\ \citenamefont {{Fisher}}}]{SurfSci.547.324}%
  \BibitemOpen
  \bibfield  {author} {\bibinfo {author} {\bibfnamefont {F.~Y.}\ \bibnamefont
  {{Naumkin}}}, \bibinfo {author} {\bibfnamefont {J.~C.}\ \bibnamefont
  {{Polanyi}}}, \bibinfo {author} {\bibfnamefont {D.}~\bibnamefont {{Rogers}}},
  \bibinfo {author} {\bibfnamefont {W.}~\bibnamefont {{Hofer}}}, \ and\
  \bibinfo {author} {\bibfnamefont {A.}~\bibnamefont {{Fisher}}},\ }\href
  {\doibase 10.1016/j.susc.2003.09.042} {\bibfield  {journal} {\bibinfo
  {journal} {Surf. Sci.}\ }\textbf {\bibinfo {volume} {547}},\ \bibinfo {pages}
  {324} (\bibinfo {year} {2003})}\BibitemShut {NoStop}%
\bibitem [{\citenamefont {{Shimomura}}\ \emph {et~al.}(2003)\citenamefont
  {{Shimomura}}, \citenamefont {{Munakata}}, \citenamefont {{Honma}},
  \citenamefont {{Widstrand}}, \citenamefont {{Johansson}}, \citenamefont
  {{Abukawa}},\ and\ \citenamefont {{Kono}}}]{SurfRevLett.10.499}%
  \BibitemOpen
  \bibfield  {author} {\bibinfo {author} {\bibfnamefont {M.}~\bibnamefont
  {{Shimomura}}}, \bibinfo {author} {\bibfnamefont {M.}~\bibnamefont
  {{Munakata}}}, \bibinfo {author} {\bibfnamefont {K.}~\bibnamefont {{Honma}}},
  \bibinfo {author} {\bibfnamefont {S.~M.}\ \bibnamefont {{Widstrand}}},
  \bibinfo {author} {\bibfnamefont {L.}~\bibnamefont {{Johansson}}}, \bibinfo
  {author} {\bibfnamefont {T.}~\bibnamefont {{Abukawa}}}, \ and\ \bibinfo
  {author} {\bibfnamefont {S.}~\bibnamefont {{Kono}}},\ }\href {\doibase
  10.1142/S0218625X03005013} {\bibfield  {journal} {\bibinfo  {journal} {Surf.
  Rev. Lett.}\ }\textbf {\bibinfo {volume} {10}},\ \bibinfo {pages} {499}
  (\bibinfo {year} {2003})}\BibitemShut {NoStop}%
\bibitem [{\citenamefont {Kim}\ \emph {et~al.}(2005)\citenamefont {Kim},
  \citenamefont {Lee},\ and\ \citenamefont {Yeom}}]{PhysRevB.71.115311}%
  \BibitemOpen
  \bibfield  {author} {\bibinfo {author} {\bibfnamefont {Y.~K.}\ \bibnamefont
  {Kim}}, \bibinfo {author} {\bibfnamefont {M.~H.}\ \bibnamefont {Lee}}, \ and\
  \bibinfo {author} {\bibfnamefont {H.~W.}\ \bibnamefont {Yeom}},\ }\href
  {\doibase 10.1103/PhysRevB.71.115311} {\bibfield  {journal} {\bibinfo
  {journal} {Phys. Rev. B}\ }\textbf {\bibinfo {volume} {71}},\ \bibinfo
  {pages} {115311} (\bibinfo {year} {2005})}\BibitemShut {NoStop}%
\bibitem [{\citenamefont {Witkowski}\ \emph {et~al.}(2005)\citenamefont
  {Witkowski}, \citenamefont {Pluchery},\ and\ \citenamefont
  {Borensztein}}]{PhysRevB.72.075354}%
  \BibitemOpen
  \bibfield  {author} {\bibinfo {author} {\bibfnamefont {N.}~\bibnamefont
  {Witkowski}}, \bibinfo {author} {\bibfnamefont {O.}~\bibnamefont {Pluchery}},
  \ and\ \bibinfo {author} {\bibfnamefont {Y.}~\bibnamefont {Borensztein}},\
  }\href {\doibase 10.1103/PhysRevB.72.075354} {\bibfield  {journal} {\bibinfo
  {journal} {Phys. Rev. B}\ }\textbf {\bibinfo {volume} {72}},\ \bibinfo
  {pages} {075354} (\bibinfo {year} {2005})}\BibitemShut {NoStop}%
\bibitem [{\citenamefont {{Nisbet}}\ \emph {et~al.}(2008)\citenamefont
  {{Nisbet}}, \citenamefont {{Lamont}}, \citenamefont {{Polcik}}, \citenamefont
  {{Terborg}}, \citenamefont {{Sayago}}, \citenamefont {{Kittel}},
  \citenamefont {{Hoeft}}, \citenamefont {{Toomes}},\ and\ \citenamefont
  {{Woodruff}}}]{JPhysCondMatt.20.304206}%
  \BibitemOpen
  \bibfield  {author} {\bibinfo {author} {\bibfnamefont {G.}~\bibnamefont
  {{Nisbet}}}, \bibinfo {author} {\bibfnamefont {C.~L.~A.}\ \bibnamefont
  {{Lamont}}}, \bibinfo {author} {\bibfnamefont {M.}~\bibnamefont {{Polcik}}},
  \bibinfo {author} {\bibfnamefont {R.}~\bibnamefont {{Terborg}}}, \bibinfo
  {author} {\bibfnamefont {D.~I.}\ \bibnamefont {{Sayago}}}, \bibinfo {author}
  {\bibfnamefont {M.}~\bibnamefont {{Kittel}}}, \bibinfo {author}
  {\bibfnamefont {J.~T.}\ \bibnamefont {{Hoeft}}}, \bibinfo {author}
  {\bibfnamefont {R.~L.}\ \bibnamefont {{Toomes}}}, \ and\ \bibinfo {author}
  {\bibfnamefont {D.~P.}\ \bibnamefont {{Woodruff}}},\ }\href {\doibase
  10.1088/0953-8984/20/30/304206} {\bibfield  {journal} {\bibinfo  {journal}
  {J. Phys.: Condens. Matter}\ }\textbf {\bibinfo {volume} {20}},\ \bibinfo
  {eid} {304206} (\bibinfo {year} {2008})}\BibitemShut {NoStop}%
\bibitem [{\citenamefont {Naydenov}\ and\ \citenamefont
  {Widdra}(2007)}]{JChemPhys.127.154711}%
  \BibitemOpen
  \bibfield  {author} {\bibinfo {author} {\bibfnamefont {B.}~\bibnamefont
  {Naydenov}}\ and\ \bibinfo {author} {\bibfnamefont {W.}~\bibnamefont
  {Widdra}},\ }\href@noop {} {\bibfield  {journal} {\bibinfo  {journal} {J.
  Chem. Phys.}\ }\textbf {\bibinfo {volume} {127}},\ \bibinfo {eid} {154711}
  (\bibinfo {year} {2007})}\BibitemShut {NoStop}%
\bibitem [{\citenamefont {{Craig}}(1993)}]{SurfSciLett.280.L279}%
  \BibitemOpen
  \bibfield  {author} {\bibinfo {author} {\bibfnamefont {B.~I.}\ \bibnamefont
  {{Craig}}},\ }\href {\doibase 10.1016/0039-6028(93)90675-A} {\bibfield
  {journal} {\bibinfo  {journal} {Surf. Sci.}\ }\textbf {\bibinfo {volume}
  {280}},\ \bibinfo {pages} {L279} (\bibinfo {year} {1993})}\BibitemShut
  {NoStop}%
\bibitem [{\citenamefont {{Jeong}}\ \emph {et~al.}(1995)\citenamefont
  {{Jeong}}, \citenamefont {{Ryu}}, \citenamefont {{Lee}},\ and\ \citenamefont
  {{Kim}}}]{SurfSci.344.L1226}%
  \BibitemOpen
  \bibfield  {author} {\bibinfo {author} {\bibfnamefont {H.~D.}\ \bibnamefont
  {{Jeong}}}, \bibinfo {author} {\bibfnamefont {S.}~\bibnamefont {{Ryu}}},
  \bibinfo {author} {\bibfnamefont {Y.~S.}\ \bibnamefont {{Lee}}}, \ and\
  \bibinfo {author} {\bibfnamefont {S.}~\bibnamefont {{Kim}}},\ }\href
  {\doibase 10.1016/0039-6028(95)00931-0} {\bibfield  {journal} {\bibinfo
  {journal} {Surf. Sci.}\ }\textbf {\bibinfo {volume} {344}},\ \bibinfo {pages}
  {L1226} (\bibinfo {year} {1995})}\BibitemShut {NoStop}%
\bibitem [{\citenamefont {{Birkenheuer}}\ \emph {et~al.}(1998)\citenamefont
  {{Birkenheuer}}, \citenamefont {{Gutdeutsch}},\ and\ \citenamefont
  {{R{\"o}sch}}}]{SurfSci.409.213}%
  \BibitemOpen
  \bibfield  {author} {\bibinfo {author} {\bibfnamefont {U.}~\bibnamefont
  {{Birkenheuer}}}, \bibinfo {author} {\bibfnamefont {U.}~\bibnamefont
  {{Gutdeutsch}}}, \ and\ \bibinfo {author} {\bibfnamefont {N.}~\bibnamefont
  {{R{\"o}sch}}},\ }\href {\doibase 10.1016/S0039-6028(98)00207-6} {\bibfield
  {journal} {\bibinfo  {journal} {Surf. Sci.}\ }\textbf {\bibinfo {volume}
  {409}},\ \bibinfo {pages} {213} (\bibinfo {year} {1998})}\BibitemShut
  {NoStop}%
\bibitem [{\citenamefont {{Wolkow}}\ \emph {et~al.}(1998)\citenamefont
  {{Wolkow}}, \citenamefont {{Lopinski}},\ and\ \citenamefont
  {{Moffatt}}}]{SurfSci.416.L1107}%
  \BibitemOpen
  \bibfield  {author} {\bibinfo {author} {\bibfnamefont {R.~A.}\ \bibnamefont
  {{Wolkow}}}, \bibinfo {author} {\bibfnamefont {G.~P.}\ \bibnamefont
  {{Lopinski}}}, \ and\ \bibinfo {author} {\bibfnamefont {D.~J.}\ \bibnamefont
  {{Moffatt}}},\ }\href {\doibase 10.1016/S0039-6028(98)00629-3} {\bibfield
  {journal} {\bibinfo  {journal} {Surf. Sci.}\ }\textbf {\bibinfo {volume}
  {416}},\ \bibinfo {pages} {L1107} (\bibinfo {year} {1998})}\BibitemShut
  {NoStop}%
\bibitem [{\citenamefont {{Kone{\v c}n{\'y}}}\ and\ \citenamefont
  {{Doren}}(1998)}]{SurfSci.417.169}%
  \BibitemOpen
  \bibfield  {author} {\bibinfo {author} {\bibfnamefont {R.}~\bibnamefont
  {{Kone{\v c}n{\'y}}}}\ and\ \bibinfo {author} {\bibfnamefont {D.~J.}\
  \bibnamefont {{Doren}}},\ }\href {\doibase 10.1016/S0039-6028(98)00554-8}
  {\bibfield  {journal} {\bibinfo  {journal} {Surf. Sci.}\ }\textbf {\bibinfo
  {volume} {417}},\ \bibinfo {pages} {169} (\bibinfo {year}
  {1998})}\BibitemShut {NoStop}%
\bibitem [{\citenamefont {Silvestrelli}\ \emph {et~al.}(2000)\citenamefont
  {Silvestrelli}, \citenamefont {Ancilotto},\ and\ \citenamefont
  {Toigo}}]{PhysRevB.62.1596}%
  \BibitemOpen
  \bibfield  {author} {\bibinfo {author} {\bibfnamefont {P.~L.}\ \bibnamefont
  {Silvestrelli}}, \bibinfo {author} {\bibfnamefont {F.}~\bibnamefont
  {Ancilotto}}, \ and\ \bibinfo {author} {\bibfnamefont {F.}~\bibnamefont
  {Toigo}},\ }\href {\doibase 10.1103/PhysRevB.62.1596} {\bibfield  {journal}
  {\bibinfo  {journal} {Phys. Rev. B}\ }\textbf {\bibinfo {volume} {62}},\
  \bibinfo {pages} {1596} (\bibinfo {year} {2000})}\BibitemShut {NoStop}%
\bibitem [{\citenamefont {{Hofer}}\ \emph {et~al.}(2001)\citenamefont
  {{Hofer}}, \citenamefont {{Fisher}}, \citenamefont {{Lopinski}},\ and\
  \citenamefont {{Wolkow}}}]{SurfSci.482.1181}%
  \BibitemOpen
  \bibfield  {author} {\bibinfo {author} {\bibfnamefont {W.~A.}\ \bibnamefont
  {{Hofer}}}, \bibinfo {author} {\bibfnamefont {A.~J.}\ \bibnamefont
  {{Fisher}}}, \bibinfo {author} {\bibfnamefont {G.~P.}\ \bibnamefont
  {{Lopinski}}}, \ and\ \bibinfo {author} {\bibfnamefont {R.~A.}\ \bibnamefont
  {{Wolkow}}},\ }\href {\doibase 10.1016/S0039-6028(01)00941-4} {\bibfield
  {journal} {\bibinfo  {journal} {Surf. Sci.}\ }\textbf {\bibinfo {volume}
  {482}},\ \bibinfo {pages} {1181} (\bibinfo {year} {2001})}\BibitemShut
  {NoStop}%
\bibitem [{\citenamefont {Hofer}\ \emph {et~al.}(2001)\citenamefont {Hofer},
  \citenamefont {Fisher}, \citenamefont {Lopinski},\ and\ \citenamefont
  {Wolkow}}]{PhysRevB.63.085314}%
  \BibitemOpen
  \bibfield  {author} {\bibinfo {author} {\bibfnamefont {W.~A.}\ \bibnamefont
  {Hofer}}, \bibinfo {author} {\bibfnamefont {A.~J.}\ \bibnamefont {Fisher}},
  \bibinfo {author} {\bibfnamefont {G.~P.}\ \bibnamefont {Lopinski}}, \ and\
  \bibinfo {author} {\bibfnamefont {R.~A.}\ \bibnamefont {Wolkow}},\ }\href
  {\doibase 10.1103/PhysRevB.63.085314} {\bibfield  {journal} {\bibinfo
  {journal} {Phys. Rev. B}\ }\textbf {\bibinfo {volume} {63}},\ \bibinfo
  {pages} {085314} (\bibinfo {year} {2001})}\BibitemShut {NoStop}%
\bibitem [{\citenamefont {{Jung}}\ and\ \citenamefont
  {{Gordon}}(2005)}]{JAmChemSoc.127.3131}%
  \BibitemOpen
  \bibfield  {author} {\bibinfo {author} {\bibfnamefont {J.}~\bibnamefont
  {{Jung}}}\ and\ \bibinfo {author} {\bibfnamefont {M.~S.}\ \bibnamefont
  {{Gordon}}},\ }\href@noop {} {\bibfield  {journal} {\bibinfo  {journal} {J.
  Am. Chem. Soc.}\ }\textbf {\bibinfo {volume} {127}},\ \bibinfo {pages} {3131}
  (\bibinfo {year} {2005})}\BibitemShut {NoStop}%
\bibitem [{\citenamefont {Lee}\ and\ \citenamefont
  {Cho}(2005)}]{PhysRevB.72.235317}%
  \BibitemOpen
  \bibfield  {author} {\bibinfo {author} {\bibfnamefont {J.-Y.}\ \bibnamefont
  {Lee}}\ and\ \bibinfo {author} {\bibfnamefont {J.-H.}\ \bibnamefont {Cho}},\
  }\href {\doibase 10.1103/PhysRevB.72.235317} {\bibfield  {journal} {\bibinfo
  {journal} {Phys. Rev. B}\ }\textbf {\bibinfo {volume} {72}},\ \bibinfo
  {pages} {235317} (\bibinfo {year} {2005})}\BibitemShut {NoStop}%
\bibitem [{\citenamefont {Mamatkulov}\ \emph {et~al.}(2006)\citenamefont
  {Mamatkulov}, \citenamefont {Stauffer}, \citenamefont {Minot},\ and\
  \citenamefont {Sonnet}}]{PhysRevB.73.035321}%
  \BibitemOpen
  \bibfield  {author} {\bibinfo {author} {\bibfnamefont {M.}~\bibnamefont
  {Mamatkulov}}, \bibinfo {author} {\bibfnamefont {L.}~\bibnamefont
  {Stauffer}}, \bibinfo {author} {\bibfnamefont {C.}~\bibnamefont {Minot}}, \
  and\ \bibinfo {author} {\bibfnamefont {P.}~\bibnamefont {Sonnet}},\ }\href
  {\doibase 10.1103/PhysRevB.73.035321} {\bibfield  {journal} {\bibinfo
  {journal} {Phys. Rev. B}\ }\textbf {\bibinfo {volume} {73}},\ \bibinfo
  {pages} {035321} (\bibinfo {year} {2006})}\BibitemShut {NoStop}%
\bibitem [{\citenamefont {Johnston}\ and\ \citenamefont
  {Nieminen}(2007)}]{PhysRevB.76.085402}%
  \BibitemOpen
  \bibfield  {author} {\bibinfo {author} {\bibfnamefont {K.}~\bibnamefont
  {Johnston}}\ and\ \bibinfo {author} {\bibfnamefont {R.~M.}\ \bibnamefont
  {Nieminen}},\ }\href {\doibase 10.1103/PhysRevB.76.085402} {\bibfield
  {journal} {\bibinfo  {journal} {Phys. Rev. B}\ }\textbf {\bibinfo {volume}
  {76}},\ \bibinfo {pages} {085402} (\bibinfo {year} {2007})}\BibitemShut
  {NoStop}%
\bibitem [{\citenamefont {Johnston}\ \emph
  {et~al.}(2008{\natexlab{a}})\citenamefont {Johnston}, \citenamefont {Kleis},
  \citenamefont {Lundqvist},\ and\ \citenamefont
  {Nieminen}}]{PhysRevB.77.121404}%
  \BibitemOpen
  \bibfield  {author} {\bibinfo {author} {\bibfnamefont {K.}~\bibnamefont
  {Johnston}}, \bibinfo {author} {\bibfnamefont {J.}~\bibnamefont {Kleis}},
  \bibinfo {author} {\bibfnamefont {B.~I.}\ \bibnamefont {Lundqvist}}, \ and\
  \bibinfo {author} {\bibfnamefont {R.~M.}\ \bibnamefont {Nieminen}},\ }\href
  {\doibase 10.1103/PhysRevB.77.121404} {\bibfield  {journal} {\bibinfo
  {journal} {Phys. Rev. B}\ }\textbf {\bibinfo {volume} {77}},\ \bibinfo
  {pages} {121404(R)} (\bibinfo {year} {2008}{\natexlab{a}}); {\it ibid.}
  \textbf {\bibinfo {volume} {77}},\ \bibinfo
  {pages} {209904(E)} (\bibinfo {year} {2008}{\natexlab{b}})}\BibitemShut
  {NoStop}%
\bibitem [{\citenamefont {Kim}\ \emph {et~al.}(2012)\citenamefont {Kim},
  \citenamefont {Tkatchenko}, \citenamefont {Cho},\ and\ \citenamefont
  {Scheffler}}]{PhysRevB.85.041403}%
  \BibitemOpen
  \bibfield  {author} {\bibinfo {author} {\bibfnamefont {H.-J.}\ \bibnamefont
  {Kim}}, \bibinfo {author} {\bibfnamefont {A.}~\bibnamefont {Tkatchenko}},
  \bibinfo {author} {\bibfnamefont {J.-H.}\ \bibnamefont {Cho}}, \ and\
  \bibinfo {author} {\bibfnamefont {M.}~\bibnamefont {Scheffler}},\ }\href
  {\doibase 10.1103/PhysRevB.85.041403} {\bibfield  {journal} {\bibinfo
  {journal} {Phys. Rev. B}\ }\textbf {\bibinfo {volume} {85}},\ \bibinfo
  {pages} {041403(R)} (\bibinfo {year} {2012})}\BibitemShut {NoStop}%
\bibitem{Note1}PBE+vdW tends to overestimate the adsorption energies,
  and we expect that PBE+vdW$^{\rm surf}$ of Ref.~\onlinecite{PhysRevLett.108.146103}
  provides more accurate adsorption energies. See also Refs.~\onlinecite{PhysRevB.86.245405}
  and \onlinecite{PhysRevB.93.035118}.
\bibitem [{\citenamefont {Czekala}\ \emph {et~al.}(2014)\citenamefont
  {Czekala}, \citenamefont {Panosetti}, \citenamefont {Lin},\ and\
  \citenamefont {Hofer}}]{SurfSci.621.152}%
  \BibitemOpen
  \bibfield  {author} {\bibinfo {author} {\bibfnamefont {P.~T.}\ \bibnamefont
  {Czekala}}, \bibinfo {author} {\bibfnamefont {C.}~\bibnamefont {Panosetti}},
  \bibinfo {author} {\bibfnamefont {H.}~\bibnamefont {Lin}}, \ and\ \bibinfo
  {author} {\bibfnamefont {W.~A.}\ \bibnamefont {Hofer}},\ }\href@noop {}
  {\bibfield  {journal} {\bibinfo  {journal} {Surf. Sci.}\ }\textbf {\bibinfo
  {volume} {621}},\ \bibinfo {pages} {152 } (\bibinfo {year}
  {2014})}\BibitemShut {NoStop}%
\bibitem [{\citenamefont {M\o{}ller}\ and\ \citenamefont
  {Plesset}(1934)}]{PhysRev.46.618}%
  \BibitemOpen
  \bibfield  {author} {\bibinfo {author} {\bibfnamefont {C.}~\bibnamefont
  {M\o{}ller}}\ and\ \bibinfo {author} {\bibfnamefont {M.~S.}\ \bibnamefont
  {Plesset}},\ }\href {\doibase 10.1103/PhysRev.46.618} {\bibfield  {journal}
  {\bibinfo  {journal} {Phys. Rev.}\ }\textbf {\bibinfo {volume} {46}},\
  \bibinfo {pages} {618} (\bibinfo {year} {1934})}\BibitemShut {NoStop}%
\bibitem [{\citenamefont {Dion}\ \emph {et~al.}(2004)\citenamefont {Dion},
  \citenamefont {Rydberg}, \citenamefont {Schr\"oder}, \citenamefont
  {Langreth},\ and\ \citenamefont {Lundqvist}}]{PhysRevLett.92.246401}%
  \BibitemOpen
  \bibfield  {author} {\bibinfo {author} {\bibfnamefont {M.}~\bibnamefont
  {Dion}}, \bibinfo {author} {\bibfnamefont {H.}~\bibnamefont {Rydberg}},
  \bibinfo {author} {\bibfnamefont {E.}~\bibnamefont {Schr\"oder}}, \bibinfo
  {author} {\bibfnamefont {D.~C.}\ \bibnamefont {Langreth}}, \ and\ \bibinfo
  {author} {\bibfnamefont {B.~I.}\ \bibnamefont {Lundqvist}},\ }\href {\doibase
  10.1103/PhysRevLett.92.246401} {\bibfield  {journal} {\bibinfo  {journal}
  {Phys. Rev. Lett.}\ }\textbf {\bibinfo {volume} {92}},\ \bibinfo {pages}
          {246401} (\bibinfo {year} {2004}); {\it ibid.} \textbf {\bibinfo {volume} {92}},
          \bibinfo {pages} {109902(E)} (\bibinfo {year} {2005})} \BibitemShut {NoStop}%
\bibitem [{\citenamefont {Langreth}\ \emph {et~al.}(2009)\citenamefont
  {Langreth}, \citenamefont {Lundqvist}, \citenamefont {Chakarova-K\"{a}ck},
  \citenamefont {Cooper}, \citenamefont {Dion}, \citenamefont {Hyldgaard},
  \citenamefont {Kelkkanen}, \citenamefont {Kleis}, \citenamefont {Kong},
  \citenamefont {Li}, \citenamefont {Moses}, \citenamefont {Murray},
  \citenamefont {Puzder}, \citenamefont {Rydberg}, \citenamefont
  {Schr\"{o}der},\ and\ \citenamefont {Thonhauser}}]{JPhysCondMat.21.084203}%
  \BibitemOpen
  \bibfield  {author} {\bibinfo {author} {\bibfnamefont {D.~C.}\ \bibnamefont
  {Langreth}}, \bibinfo {author} {\bibfnamefont {B.~I.}\ \bibnamefont
  {Lundqvist}}, \bibinfo {author} {\bibfnamefont {S.~D.}\ \bibnamefont
  {Chakarova-K\"{a}ck}}, \bibinfo {author} {\bibfnamefont {V.~R.}\ \bibnamefont
  {Cooper}}, \bibinfo {author} {\bibfnamefont {M.}~\bibnamefont {Dion}},
  \bibinfo {author} {\bibfnamefont {P.}~\bibnamefont {Hyldgaard}}, \bibinfo
  {author} {\bibfnamefont {A.}~\bibnamefont {Kelkkanen}}, \bibinfo {author}
  {\bibfnamefont {J.}~\bibnamefont {Kleis}}, \bibinfo {author} {\bibfnamefont
  {L.}~\bibnamefont {Kong}}, \bibinfo {author} {\bibfnamefont {S.}~\bibnamefont
  {Li}}, \bibinfo {author} {\bibfnamefont {P.~G.}\ \bibnamefont {Moses}},
  \bibinfo {author} {\bibfnamefont {E.}~\bibnamefont {Murray}}, \bibinfo
  {author} {\bibfnamefont {A.}~\bibnamefont {Puzder}}, \bibinfo {author}
  {\bibfnamefont {H.}~\bibnamefont {Rydberg}}, \bibinfo {author} {\bibfnamefont
  {E.}~\bibnamefont {Schr\"{o}der}}, \ and\ \bibinfo {author} {\bibfnamefont
  {T.}~\bibnamefont {Thonhauser}},\ }\href@noop {} {\bibfield  {journal}
  {\bibinfo  {journal} {J. Phys.: Condens. Matter}\ }\textbf {\bibinfo {volume}
  {21}},\ \bibinfo {pages} {084203} (\bibinfo {year} {2009})}\BibitemShut
  {NoStop}%
\bibitem [{\citenamefont {Toyoda}\ \emph {et~al.}(2010)\citenamefont {Toyoda},
  \citenamefont {Hamada}, \citenamefont {Lee}, \citenamefont {Yanagisawa},\
  and\ \citenamefont {Morikawa}}]{JChemPhys.132.134703}%
  \BibitemOpen
  \bibfield  {author} {\bibinfo {author} {\bibfnamefont {K.}~\bibnamefont
  {Toyoda}}, \bibinfo {author} {\bibfnamefont {I.}~\bibnamefont {Hamada}},
  \bibinfo {author} {\bibfnamefont {K.}~\bibnamefont {Lee}}, \bibinfo {author}
  {\bibfnamefont {S.}~\bibnamefont {Yanagisawa}}, \ and\ \bibinfo {author}
  {\bibfnamefont {Y.}~\bibnamefont {Morikawa}},\ }\href@noop {} {\bibfield
  {journal} {\bibinfo  {journal} {J. Chem. Phys.}\ }\textbf {\bibinfo {volume}
  {132}},\ \bibinfo {eid} {134703} (\bibinfo {year} {2010})}\BibitemShut
  {NoStop}%
\bibitem [{\citenamefont {Rom\'an-P\'erez}\ and\ \citenamefont
  {Soler}(2009)}]{PhysRevLett.103.096102}%
  \BibitemOpen
  \bibfield  {author} {\bibinfo {author} {\bibfnamefont {G.}~\bibnamefont
  {Rom\'an-P\'erez}}\ and\ \bibinfo {author} {\bibfnamefont {J.~M.}\
  \bibnamefont {Soler}},\ }\href {\doibase 10.1103/PhysRevLett.103.096102}
  {\bibfield  {journal} {\bibinfo  {journal} {Phys. Rev. Lett.}\ }\textbf
  {\bibinfo {volume} {103}},\ \bibinfo {pages} {096102} (\bibinfo {year}
  {2009})}\BibitemShut {NoStop}%
\bibitem [{\citenamefont {Gulans}\ \emph {et~al.}(2009)\citenamefont {Gulans},
  \citenamefont {Puska},\ and\ \citenamefont {Nieminen}}]{PhysRevB.79.201105}%
  \BibitemOpen
  \bibfield  {author} {\bibinfo {author} {\bibfnamefont {A.}~\bibnamefont
  {Gulans}}, \bibinfo {author} {\bibfnamefont {M.~J.}\ \bibnamefont {Puska}}, \
  and\ \bibinfo {author} {\bibfnamefont {R.~M.}\ \bibnamefont {Nieminen}},\
  }\href {\doibase 10.1103/PhysRevB.79.201105} {\bibfield  {journal} {\bibinfo
  {journal} {Phys. Rev. B}\ }\textbf {\bibinfo {volume} {79}},\ \bibinfo
  {pages} {201105} (\bibinfo {year} {2009})}\BibitemShut {NoStop}%
\bibitem [{\citenamefont {{Wu}}\ and\ \citenamefont
  {{Gygi}}(2012)}]{JChemPhys.136.224107}%
  \BibitemOpen
  \bibfield  {author} {\bibinfo {author} {\bibfnamefont {J.}~\bibnamefont
  {{Wu}}}\ and\ \bibinfo {author} {\bibfnamefont {F.}~\bibnamefont {{Gygi}}},\
  }\href {\doibase 10.1063/1.4727850} {\bibfield  {journal} {\bibinfo
  {journal} {J. Chem. Phys.}\ }\textbf {\bibinfo {volume} {136}},\ \bibinfo
  {pages} {224107} (\bibinfo {year} {2012})}\BibitemShut {NoStop}%
\bibitem [{\citenamefont {Cooper}(2010)}]{PhysRevB.81.161104}%
  \BibitemOpen
  \bibfield  {author} {\bibinfo {author} {\bibfnamefont {V.~R.}\ \bibnamefont
  {Cooper}},\ }\href {\doibase 10.1103/PhysRevB.81.161104} {\bibfield
  {journal} {\bibinfo  {journal} {Phys. Rev. B}\ }\textbf {\bibinfo {volume}
  {81}},\ \bibinfo {pages} {161104(R)} (\bibinfo {year} {2010})}\BibitemShut
  {NoStop}%
\bibitem [{\citenamefont {Lee}\ \emph {et~al.}(2010)\citenamefont {Lee},
  \citenamefont {Murray}, \citenamefont {Kong}, \citenamefont {Lundqvist},\
  and\ \citenamefont {Langreth}}]{PhysRevB.82.081101}%
  \BibitemOpen
  \bibfield  {author} {\bibinfo {author} {\bibfnamefont {K.}~\bibnamefont
  {Lee}}, \bibinfo {author} {\bibfnamefont {{\'E}.~D.}\ \bibnamefont {Murray}},
  \bibinfo {author} {\bibfnamefont {L.}~\bibnamefont {Kong}}, \bibinfo {author}
  {\bibfnamefont {B.~I.}\ \bibnamefont {Lundqvist}}, \ and\ \bibinfo {author}
  {\bibfnamefont {D.~C.}\ \bibnamefont {Langreth}},\ }\href {\doibase
  10.1103/PhysRevB.82.081101} {\bibfield  {journal} {\bibinfo  {journal} {Phys.
  Rev. B}\ }\textbf {\bibinfo {volume} {82}},\ \bibinfo {pages} {081101(R)}
  (\bibinfo {year} {2010})}\BibitemShut {NoStop}%
\bibitem [{\citenamefont {{Klime{\v s}}}\ \emph {et~al.}(2010)\citenamefont
  {{Klime{\v s}}}, \citenamefont {{Bowler}},\ and\ \citenamefont
  {{Michaelides}}}]{JPhysCondMatt.22.022201}%
  \BibitemOpen
  \bibfield  {author} {\bibinfo {author} {\bibfnamefont {J.}~\bibnamefont
  {{Klime{\v s}}}}, \bibinfo {author} {\bibfnamefont {D.~R.}\ \bibnamefont
  {{Bowler}}}, \ and\ \bibinfo {author} {\bibfnamefont {A.}~\bibnamefont
  {{Michaelides}}},\ }\href {\doibase 10.1088/0953-8984/22/2/022201} {\bibfield
   {journal} {\bibinfo  {journal} {J. Phys.: Condens. Matter}\ }\textbf
  {\bibinfo {volume} {22}},\ \bibinfo {eid} {022201} (\bibinfo {year}
  {2010})}\BibitemShut {NoStop}%
\bibitem [{\citenamefont {{Klime{\v s}}}\ \emph {et~al.}(2011)\citenamefont
  {{Klime{\v s}}}, \citenamefont {{Bowler}},\ and\ \citenamefont
  {{Michaelides}}}]{PhysRevB.83.195131}%
  \BibitemOpen
  \bibfield  {author} {\bibinfo {author} {\bibfnamefont {J.}~\bibnamefont
  {{Klime{\v s}}}}, \bibinfo {author} {\bibfnamefont {D.~R.}\ \bibnamefont
  {{Bowler}}}, \ and\ \bibinfo {author} {\bibfnamefont {A.}~\bibnamefont
  {{Michaelides}}},\ }\href {\doibase 10.1103/PhysRevB.83.195131} {\bibfield
  {journal} {\bibinfo  {journal} {Phys. Rev. B}\ }\textbf {\bibinfo {volume}
  {83}},\ \bibinfo {eid} {195131} (\bibinfo {year} {2011})}\BibitemShut
  {NoStop}%
\bibitem [{\citenamefont {Wellendorff}\ and\ \citenamefont
  {Bligaard}(2011)}]{TopCatal.54.1143}%
  \BibitemOpen
  \bibfield  {author} {\bibinfo {author} {\bibfnamefont {J.}~\bibnamefont
  {Wellendorff}}\ and\ \bibinfo {author} {\bibfnamefont {T.}~\bibnamefont
  {Bligaard}},\ }\href {\doibase 10.1007/s11244-011-9736-4} {\bibfield
  {journal} {\bibinfo  {journal} {Top. Catal.}\ }\textbf {\bibinfo {volume}
  {54}},\ \bibinfo {pages} {1143} (\bibinfo {year} {2011})}\BibitemShut
  {NoStop}%
\bibitem [{\citenamefont {{Wellendorff}}\ \emph {et~al.}(2012)\citenamefont
  {{Wellendorff}}, \citenamefont {{Lundgaard}}, \citenamefont
  {{M{\o}gelh{\o}j}}, \citenamefont {{Petzold}}, \citenamefont {{Landis}},
  \citenamefont {{N{\o}rskov}}, \citenamefont {{Bligaard}},\ and\ \citenamefont
  {{Jacobsen}}}]{PhysRevB.85.235149}%
  \BibitemOpen
  \bibfield  {author} {\bibinfo {author} {\bibfnamefont {J.}~\bibnamefont
  {{Wellendorff}}}, \bibinfo {author} {\bibfnamefont {K.~T.}\ \bibnamefont
  {{Lundgaard}}}, \bibinfo {author} {\bibfnamefont {A.}~\bibnamefont
  {{M{\o}gelh{\o}j}}}, \bibinfo {author} {\bibfnamefont {V.}~\bibnamefont
  {{Petzold}}}, \bibinfo {author} {\bibfnamefont {D.~D.}\ \bibnamefont
  {{Landis}}}, \bibinfo {author} {\bibfnamefont {J.~K.}\ \bibnamefont
  {{N{\o}rskov}}}, \bibinfo {author} {\bibfnamefont {T.}~\bibnamefont
  {{Bligaard}}}, \ and\ \bibinfo {author} {\bibfnamefont {K.~W.}\ \bibnamefont
  {{Jacobsen}}},\ }\href {\doibase 10.1103/PhysRevB.85.235149} {\bibfield
  {journal} {\bibinfo  {journal} {Phys. Rev. B}\ }\textbf {\bibinfo {volume}
  {85}},\ \bibinfo {eid} {235149} (\bibinfo {year} {2012})}\BibitemShut
  {NoStop}%
\bibitem [{\citenamefont {{Berland}}\ and\ \citenamefont
  {{Hyldgaard}}(2014)}]{PhysRevB.89.035412}%
  \BibitemOpen
  \bibfield  {author} {\bibinfo {author} {\bibfnamefont {K.}~\bibnamefont
  {{Berland}}}\ and\ \bibinfo {author} {\bibfnamefont {P.}~\bibnamefont
  {{Hyldgaard}}},\ }\href {\doibase 10.1103/PhysRevB.89.035412} {\bibfield
  {journal} {\bibinfo  {journal} {Phys. Rev. B}\ }\textbf {\bibinfo {volume}
  {89}},\ \bibinfo {eid} {035412} (\bibinfo {year} {2014})}\BibitemShut
  {NoStop}%
\bibitem [{\citenamefont {Hamada}(2014)}]{PhysRevB.89.121103}%
  \BibitemOpen
  \bibfield  {author} {\bibinfo {author} {\bibfnamefont {I.}~\bibnamefont
  {Hamada}},\ }\href {\doibase 10.1103/PhysRevB.89.121103} {\bibfield
  {journal} {\bibinfo  {journal} {Phys. Rev. B}\ }\textbf {\bibinfo {volume}
  {89}},\ \bibinfo {pages} {121103(R)} (\bibinfo {year} {2014})}\BibitemShut
  {NoStop}%
\bibitem [{\citenamefont {{Morikawa}}\ \emph {et~al.}(2001)\citenamefont
  {{Morikawa}}, \citenamefont {{Iwata}},\ and\ \citenamefont
  {{Terakura}}}]{ApplSurfSci.169.11}%
  \BibitemOpen
  \bibfield  {author} {\bibinfo {author} {\bibfnamefont {Y.}~\bibnamefont
  {{Morikawa}}}, \bibinfo {author} {\bibfnamefont {K.}~\bibnamefont {{Iwata}}},
  \ and\ \bibinfo {author} {\bibfnamefont {K.}~\bibnamefont {{Terakura}}},\
  }\href {\doibase 10.1016/S0169-4332(00)00631-0} {\bibfield  {journal}
  {\bibinfo  {journal} {Appl. Surf. Sci.}\ }\textbf {\bibinfo {volume} {169}},\
  \bibinfo {pages} {11} (\bibinfo {year} {2001})}\BibitemShut {NoStop}%
\bibitem [{\citenamefont {Troullier}\ and\ \citenamefont
  {Martins}(1991)}]{PhysRevB.43.1993}%
  \BibitemOpen
  \bibfield  {author} {\bibinfo {author} {\bibfnamefont {N.}~\bibnamefont
  {Troullier}}\ and\ \bibinfo {author} {\bibfnamefont {J.~L.}\ \bibnamefont
  {Martins}},\ }\href {\doibase 10.1103/PhysRevB.43.1993} {\bibfield  {journal}
  {\bibinfo  {journal} {Phys. Rev. B}\ }\textbf {\bibinfo {volume} {43}},\
  \bibinfo {pages} {1993} (\bibinfo {year} {1991})}\BibitemShut {NoStop}%
\bibitem [{\citenamefont {Otani}\ and\ \citenamefont
  {Sugino}(2006)}]{PhysRevB.73.115407}%
  \BibitemOpen
  \bibfield  {author} {\bibinfo {author} {\bibfnamefont {M.}~\bibnamefont
  {Otani}}\ and\ \bibinfo {author} {\bibfnamefont {O.}~\bibnamefont {Sugino}},\
  }\href {\doibase 10.1103/PhysRevB.73.115407} {\bibfield  {journal} {\bibinfo
  {journal} {Phys. Rev. B}\ }\textbf {\bibinfo {volume} {73}},\ \bibinfo
  {pages} {115407} (\bibinfo {year} {2006})}\BibitemShut {NoStop}%
\bibitem [{\citenamefont {Hamada}\ \emph {et~al.}(2009)\citenamefont {Hamada},
  \citenamefont {Otani}, \citenamefont {Sugino},\ and\ \citenamefont
  {Morikawa}}]{PhysRevB.80.165411}%
  \BibitemOpen
  \bibfield  {author} {\bibinfo {author} {\bibfnamefont {I.}~\bibnamefont
  {Hamada}}, \bibinfo {author} {\bibfnamefont {M.}~\bibnamefont {Otani}},
  \bibinfo {author} {\bibfnamefont {O.}~\bibnamefont {Sugino}}, \ and\ \bibinfo
  {author} {\bibfnamefont {Y.}~\bibnamefont {Morikawa}},\ }\href {\doibase
  10.1103/PhysRevB.80.165411} {\bibfield  {journal} {\bibinfo  {journal} {Phys.
  Rev. B}\ }\textbf {\bibinfo {volume} {80}},\ \bibinfo {pages} {165411}
  (\bibinfo {year} {2009})}\BibitemShut {NoStop}%
\bibitem [{\citenamefont {Perdew}\ \emph {et~al.}(1996)\citenamefont {Perdew},
  \citenamefont {Burke},\ and\ \citenamefont
  {Ernzerhof}}]{PhysRevLett.77.3865}
  \BibitemOpen
  \bibfield  {author} {\bibinfo {author} {\bibfnamefont {J.~P.}\ \bibnamefont
  {Perdew}}, \bibinfo {author} {\bibfnamefont {K.}~\bibnamefont {Burke}}, \
  and\ \bibinfo {author} {\bibfnamefont {M.}~\bibnamefont {Ernzerhof}},\ }\href
  {\doibase 10.1103/PhysRevLett.77.3865} {\bibfield  {journal} {\bibinfo
  {journal} {Phys. Rev. Lett.}\ }\textbf {\bibinfo {volume} {77}},\ \bibinfo
  {pages} {3865} (\bibinfo {year} {1996})}\BibitemShut {NoStop}
\bibitem [{\citenamefont {{Zhang}}\ and\ \citenamefont
  {{Yang}}(1998)}]{PhysRevLett.80.890}%
  \BibitemOpen
  \bibfield  {author} {\bibinfo {author} {\bibfnamefont {Y.}~\bibnamefont
  {{Zhang}}}\ and\ \bibinfo {author} {\bibfnamefont {W.}~\bibnamefont
  {{Yang}}},\ }\href {\doibase 10.1103/PhysRevLett.80.890} {\bibfield
  {journal} {\bibinfo  {journal} {Phys. Rev. Lett.}\ }\textbf {\bibinfo
  {volume} {80}},\ \bibinfo {pages} {890} (\bibinfo {year} {1998})}\BibitemShut
  {NoStop}%
\bibitem [{\citenamefont {Lacks}\ and\ \citenamefont
  {Gordon}(1993)}]{PhysRevA.47.4681}%
  \BibitemOpen
  \bibfield  {author} {\bibinfo {author} {\bibfnamefont {D.~J.}\ \bibnamefont
  {Lacks}}\ and\ \bibinfo {author} {\bibfnamefont {R.~G.}\ \bibnamefont
  {Gordon}},\ }\href {\doibase 10.1103/PhysRevA.47.4681} {\bibfield  {journal}
  {\bibinfo  {journal} {Phys. Rev. A}\ }\textbf {\bibinfo {volume} {47}},\
  \bibinfo {pages} {4681} (\bibinfo {year} {1993})}\BibitemShut {NoStop}%
\bibitem [{\citenamefont {{Kannemann}}\ and\ \citenamefont
  {{Becke}}(2009)}]{JChemTheoryComput.5.719}%
  \BibitemOpen
  \bibfield  {author} {\bibinfo {author} {\bibfnamefont {F.~O.}\ \bibnamefont
  {{Kannemann}}}\ and\ \bibinfo {author} {\bibfnamefont {A.~D.}\ \bibnamefont
  {{Becke}}},\ }\href@noop {} {\bibfield  {journal} {\bibinfo  {journal} {J.
  Chem. Theory Comput.}\ }\textbf {\bibinfo {volume} {5}},\ \bibinfo {pages}
  {719} (\bibinfo {year} {2009})}\BibitemShut {NoStop}%
\bibitem [{\citenamefont {{Murray}}\ \emph {et~al.}(2009)\citenamefont
  {{Murray}}, \citenamefont {{Lee}},\ and\ \citenamefont
  {{Langreth}}}]{JChemTheoryComput.5.2754}%
  \BibitemOpen
  \bibfield  {author} {\bibinfo {author} {\bibfnamefont {{\'E}.~D.}\
  \bibnamefont {{Murray}}}, \bibinfo {author} {\bibfnamefont {K.}~\bibnamefont
  {{Lee}}}, \ and\ \bibinfo {author} {\bibfnamefont {D.~C.}\ \bibnamefont
  {{Langreth}}},\ }\href@noop {} {\bibfield  {journal} {\bibinfo  {journal} {J.
  Chem. Theory Comput.}\ }\textbf {\bibinfo {volume} {5}},\ \bibinfo {pages}
  {2754} (\bibinfo {year} {2009})}\BibitemShut {NoStop}%
\bibitem [{\citenamefont {Perdew}\ and\ \citenamefont
  {Wang}(1986)}]{PhysRevB.33.8800}%
  \BibitemOpen
  \bibfield  {author} {\bibinfo {author} {\bibfnamefont {J.~P.}\ \bibnamefont
  {Perdew}}\ and\ \bibinfo {author} {\bibfnamefont {Y.}~\bibnamefont {Wang}},\
  }\href {\doibase 10.1103/PhysRevB.33.8800} {\bibfield  {journal} {\bibinfo
  {journal} {Phys. Rev. B}\ }\textbf {\bibinfo {volume} {33}},\ \bibinfo
  {pages} {8800} (\bibinfo {year} {1986})}\BibitemShut {NoStop}%
\bibitem [{\citenamefont {{Becke}}(1986)}]{JChemPhys.85.7184}%
  \BibitemOpen
  \bibfield  {author} {\bibinfo {author} {\bibfnamefont {A.~D.}\ \bibnamefont
  {{Becke}}},\ }\href {\doibase 10.1063/1.451353} {\bibfield  {journal}
  {\bibinfo  {journal} {J. Chem. Phys.}\ }\textbf {\bibinfo {volume} {85}},\
  \bibinfo {pages} {7184} (\bibinfo {year} {1986})}\BibitemShut {NoStop}%
\bibitem [{\citenamefont {{Jure{\v c}ka}}\ \emph {et~al.}(2006)\citenamefont
  {{Jure{\v c}ka}}, \citenamefont {{{\v S}poner}}, \citenamefont {{{\v
  C}ern{\'y}}},\ and\ \citenamefont {{Hobza}}}]{PhysChemChemPhys.8.1985}%
  \BibitemOpen
  \bibfield  {author} {\bibinfo {author} {\bibfnamefont {P.}~\bibnamefont
  {{Jure{\v c}ka}}}, \bibinfo {author} {\bibfnamefont {J.}~\bibnamefont {{{\v
  S}poner}}}, \bibinfo {author} {\bibfnamefont {J.}~\bibnamefont {{{\v
  C}ern{\'y}}}}, \ and\ \bibinfo {author} {\bibfnamefont {P.}~\bibnamefont
  {{Hobza}}},\ }\href {\doibase 10.1039/b600027d} {\bibfield  {journal}
  {\bibinfo  {journal} {Phys. Chem. Chem. Phys.}\ }\textbf {\bibinfo {volume}
  {8}},\ \bibinfo {pages} {1985} (\bibinfo {year} {2006})}\BibitemShut
  {NoStop}%
\bibitem [{\citenamefont {Lon\ifmmode \check{c}\else
  \v{c}\fi{}ari\ifmmode~\acute{c}\else \'{c}\fi{}}\ and\ \citenamefont
  {Despoja}(2014)}]{PhysRevB.90.075414}%
  \BibitemOpen
  \bibfield  {author} {\bibinfo {author} {\bibfnamefont {I.}~\bibnamefont
  {Lon\ifmmode \check{c}\else \v{c}\fi{}ari\ifmmode~\acute{c}\else
  \'{c}\fi{}}}\ and\ \bibinfo {author} {\bibfnamefont {V.}~\bibnamefont
  {Despoja}},\ }\href {\doibase 10.1103/PhysRevB.90.075414} {\bibfield
  {journal} {\bibinfo  {journal} {Phys. Rev. B}\ }\textbf {\bibinfo {volume}
  {90}},\ \bibinfo {pages} {075414} (\bibinfo {year} {2014})}\BibitemShut
  {NoStop}%
\bibitem [{\citenamefont {Bj{\"o}rk}\ and\ \citenamefont
  {Stafstr{\"o}m}(2014)}]{ChemPhysChem.15.2851}%
  \BibitemOpen
  \bibfield  {author} {\bibinfo {author} {\bibfnamefont {J.}~\bibnamefont
  {Bj{\"o}rk}}\ and\ \bibinfo {author} {\bibfnamefont {S.}~\bibnamefont
  {Stafstr{\"o}m}},\ }\href {\doibase 10.1002/cphc.201402063} {\bibfield
  {journal} {\bibinfo  {journal} {ChemPhysChem}\ }\textbf {\bibinfo {volume}
  {15}},\ \bibinfo {pages} {2851} (\bibinfo {year} {2014})}\BibitemShut
  {NoStop}%
\bibitem [{\citenamefont {{Han}}\ \emph {et~al.}(2014)\citenamefont {{Han}},
  \citenamefont {{Akagi}}, \citenamefont {{Canova}}, \citenamefont {{Mutoh}},
  \citenamefont {{Shiraki}}, \citenamefont {{Iwaya}}, \citenamefont {{Weiss}},
  \citenamefont {{Asao}},\ and\ \citenamefont {{Hitosugi}}}]{ACSNano.8.9181}%
  \BibitemOpen
  \bibfield  {author} {\bibinfo {author} {\bibfnamefont {P.}~\bibnamefont
  {{Han}}}, \bibinfo {author} {\bibfnamefont {K.}~\bibnamefont {{Akagi}}},
  \bibinfo {author} {\bibfnamefont {F.~F.}\ \bibnamefont {{Canova}}}, \bibinfo
  {author} {\bibfnamefont {H.}~\bibnamefont {{Mutoh}}}, \bibinfo {author}
  {\bibfnamefont {S.}~\bibnamefont {{Shiraki}}}, \bibinfo {author}
  {\bibfnamefont {K.}~\bibnamefont {{Iwaya}}}, \bibinfo {author} {\bibfnamefont
  {P.~S.}\ \bibnamefont {{Weiss}}}, \bibinfo {author} {\bibfnamefont
  {N.}~\bibnamefont {{Asao}}}, \ and\ \bibinfo {author} {\bibfnamefont
  {T.}~\bibnamefont {{Hitosugi}}},\ }\href@noop {} {\bibfield  {journal}
  {\bibinfo  {journal} {ACS Nano}\ }\textbf {\bibinfo {volume} {8}},\ \bibinfo
  {pages} {9181} (\bibinfo {year} {2014})}\BibitemShut {NoStop}%
\bibitem [{\citenamefont {Huttmann}\ \emph {et~al.}(2015)\citenamefont
  {Huttmann}, \citenamefont {Mart\'{\i}nez-Galera}, \citenamefont {Caciuc},
  \citenamefont {Atodiresei}, \citenamefont {Schumacher}, \citenamefont
  {Standop}, \citenamefont {Hamada}, \citenamefont {Wehling}, \citenamefont
  {Bl\"ugel},\ and\ \citenamefont {Michely}}]{PhysRevLett.115.236101}%
  \BibitemOpen
  \bibfield  {author} {\bibinfo {author} {\bibfnamefont {F.}~\bibnamefont
  {Huttmann}}, \bibinfo {author} {\bibfnamefont {A.~J.}\ \bibnamefont
  {Mart\'{\i}nez-Galera}}, \bibinfo {author} {\bibfnamefont {V.}~\bibnamefont
  {Caciuc}}, \bibinfo {author} {\bibfnamefont {N.}~\bibnamefont {Atodiresei}},
  \bibinfo {author} {\bibfnamefont {S.}~\bibnamefont {Schumacher}}, \bibinfo
  {author} {\bibfnamefont {S.}~\bibnamefont {Standop}}, \bibinfo {author}
  {\bibfnamefont {I.}~\bibnamefont {Hamada}}, \bibinfo {author} {\bibfnamefont
  {T.~O.}\ \bibnamefont {Wehling}}, \bibinfo {author} {\bibfnamefont
  {S.}~\bibnamefont {Bl\"ugel}}, \ and\ \bibinfo {author} {\bibfnamefont
  {T.}~\bibnamefont {Michely}},\ }\href {\doibase
  10.1103/PhysRevLett.115.236101} {\bibfield  {journal} {\bibinfo  {journal}
  {Phys. Rev. Lett.}\ }\textbf {\bibinfo {volume} {115}},\ \bibinfo {pages}
  {236101} (\bibinfo {year} {2015})}\BibitemShut {NoStop}%
\bibitem [{\citenamefont {Yang}\ \emph {et~al.}(2015)\citenamefont {Yang},
  \citenamefont {Bj\"{o}rk}, \citenamefont {Lin}, \citenamefont {Zhang},
  \citenamefont {Zhang}, \citenamefont {Li}, \citenamefont {Fan}, \citenamefont
  {Li},\ and\ \citenamefont {Chi}}]{JAmChemSoc.137.15}%
  \BibitemOpen
  \bibfield  {author} {\bibinfo {author} {\bibfnamefont {B.}~\bibnamefont
  {Yang}}, \bibinfo {author} {\bibfnamefont {J.}~\bibnamefont {Bj\"{o}rk}},
  \bibinfo {author} {\bibfnamefont {H.}~\bibnamefont {Lin}}, \bibinfo {author}
  {\bibfnamefont {X.}~\bibnamefont {Zhang}}, \bibinfo {author} {\bibfnamefont
  {H.}~\bibnamefont {Zhang}}, \bibinfo {author} {\bibfnamefont
  {Y.}~\bibnamefont {Li}}, \bibinfo {author} {\bibfnamefont {J.}~\bibnamefont
  {Fan}}, \bibinfo {author} {\bibfnamefont {Q.}~\bibnamefont {Li}}, \ and\
  \bibinfo {author} {\bibfnamefont {L.}~\bibnamefont {Chi}},\ }\href@noop {}
  {\bibfield  {journal} {\bibinfo  {journal} {J. Am. Chem. Soc.}\ }\textbf
  {\bibinfo {volume} {137}},\ \bibinfo {pages} {4904} (\bibinfo {year}
  {2015})}\BibitemShut {NoStop}%
\bibitem [{\citenamefont {Callsen}\ and\ \citenamefont
  {Hamada}(2015)}]{PhysRevB.91.195103}%
  \BibitemOpen
  \bibfield  {author} {\bibinfo {author} {\bibfnamefont {M.}~\bibnamefont
  {Callsen}}\ and\ \bibinfo {author} {\bibfnamefont {I.}~\bibnamefont
  {Hamada}},\ }\href {\doibase 10.1103/PhysRevB.91.195103} {\bibfield
  {journal} {\bibinfo  {journal} {Phys. Rev. B}\ }\textbf {\bibinfo {volume}
  {91}},\ \bibinfo {pages} {195103} (\bibinfo {year} {2015})}\BibitemShut
  {NoStop}%
\bibitem [{\citenamefont {Obata}\ \emph {et~al.}(2015)\citenamefont {Obata},
  \citenamefont {Nakamura}, \citenamefont {Hamada},\ and\ \citenamefont
  {Oda}}]{JPhysSocJpn.84.024715}%
  \BibitemOpen
  \bibfield  {author} {\bibinfo {author} {\bibfnamefont {M.}~\bibnamefont
  {Obata}}, \bibinfo {author} {\bibfnamefont {M.}~\bibnamefont {Nakamura}},
  \bibinfo {author} {\bibfnamefont {I.}~\bibnamefont {Hamada}}, \ and\ \bibinfo
  {author} {\bibfnamefont {T.}~\bibnamefont {Oda}},\ }\href {\doibase
  10.7566/JPSJ.84.024715} {\bibfield  {journal} {\bibinfo  {journal} {J. Phys.
  Soc. Jpn.}\ }\textbf {\bibinfo {volume} {84}},\ \bibinfo {pages} {024715}
  (\bibinfo {year} {2015})}\BibitemShut {NoStop}%
\bibitem [{\citenamefont {Thonhauser}\ \emph {et~al.}(2007)\citenamefont
  {Thonhauser}, \citenamefont {Cooper}, \citenamefont {Li}, \citenamefont
  {Puzder}, \citenamefont {Hyldgaard},\ and\ \citenamefont
  {Langreth}}]{PhysRevB.76.125112}%
  \BibitemOpen
  \bibfield  {author} {\bibinfo {author} {\bibfnamefont {T.}~\bibnamefont
  {Thonhauser}}, \bibinfo {author} {\bibfnamefont {V.~R.}\ \bibnamefont
  {Cooper}}, \bibinfo {author} {\bibfnamefont {S.}~\bibnamefont {Li}}, \bibinfo
  {author} {\bibfnamefont {A.}~\bibnamefont {Puzder}}, \bibinfo {author}
  {\bibfnamefont {P.}~\bibnamefont {Hyldgaard}}, \ and\ \bibinfo {author}
  {\bibfnamefont {D.~C.}\ \bibnamefont {Langreth}},\ }\href {\doibase
  10.1103/PhysRevB.76.125112} {\bibfield  {journal} {\bibinfo  {journal} {Phys.
  Rev. B}\ }\textbf {\bibinfo {volume} {76}},\ \bibinfo {pages} {125112}
  (\bibinfo {year} {2007})}\BibitemShut {NoStop}%
\bibitem [{\citenamefont {Becke}(1988)}]{PhysRevA.38.3098}%
  \BibitemOpen
  \bibfield  {author} {\bibinfo {author} {\bibfnamefont {A.~D.}\ \bibnamefont
  {Becke}},\ }\href {\doibase 10.1103/PhysRevA.38.3098} {\bibfield  {journal}
  {\bibinfo  {journal} {Phys. Rev. A}\ }\textbf {\bibinfo {volume} {38}},\
  \bibinfo {pages} {3098} (\bibinfo {year} {1988})}\BibitemShut {NoStop}%
\bibitem [{\citenamefont {Yanagisawa}\ \emph {et~al.}(2008)\citenamefont
  {Yanagisawa}, \citenamefont {Lee},\ and\ \citenamefont
  {Morikawa}}]{JChemPhys.128.244704}%
  \BibitemOpen
  \bibfield  {author} {\bibinfo {author} {\bibfnamefont {S.}~\bibnamefont
  {Yanagisawa}}, \bibinfo {author} {\bibfnamefont {K.}~\bibnamefont {Lee}}, \
  and\ \bibinfo {author} {\bibfnamefont {Y.}~\bibnamefont {Morikawa}},\
  }\href@noop {} {\bibfield  {journal} {\bibinfo  {journal} {J. Chem. Phys.}\
  }\textbf {\bibinfo {volume} {128}},\ \bibinfo {eid} {244704} (\bibinfo {year}
  {2008})}\BibitemShut {NoStop}%
\bibitem [{\citenamefont {Ruiz}\ \emph {et~al.}(2012)\citenamefont {Ruiz},
  \citenamefont {Liu}, \citenamefont {Zojer}, \citenamefont {Scheffler},\ and\
  \citenamefont {Tkatchenko}}]{PhysRevLett.108.146103}%
  \BibitemOpen
  \bibfield  {author} {\bibinfo {author} {\bibfnamefont {V.~G.}\ \bibnamefont
  {Ruiz}}, \bibinfo {author} {\bibfnamefont {W.}~\bibnamefont {Liu}}, \bibinfo
  {author} {\bibfnamefont {E.}~\bibnamefont {Zojer}}, \bibinfo {author}
  {\bibfnamefont {M.}~\bibnamefont {Scheffler}}, \ and\ \bibinfo {author}
  {\bibfnamefont {A.}~\bibnamefont {Tkatchenko}},\ }\href {\doibase
  10.1103/PhysRevLett.108.146103} {\bibfield  {journal} {\bibinfo  {journal}
  {Phys. Rev. Lett.}\ }\textbf {\bibinfo {volume} {108}},\ \bibinfo {pages}
  {146103} (\bibinfo {year} {2012})}\BibitemShut {NoStop}%
\bibitem [{\citenamefont {Liu}\ \emph {et~al.}(2012)\citenamefont {Liu},
  \citenamefont {Carrasco}, \citenamefont {Santra}, \citenamefont
  {Michaelides}, \citenamefont {Scheffler},\ and\ \citenamefont
  {Tkatchenko}}]{PhysRevB.86.245405}%
  \BibitemOpen
  \bibfield  {author} {\bibinfo {author} {\bibfnamefont {W.}~\bibnamefont
  {Liu}}, \bibinfo {author} {\bibfnamefont {J.}~\bibnamefont {Carrasco}},
  \bibinfo {author} {\bibfnamefont {B.}~\bibnamefont {Santra}}, \bibinfo
  {author} {\bibfnamefont {A.}~\bibnamefont {Michaelides}}, \bibinfo {author}
  {\bibfnamefont {M.}~\bibnamefont {Scheffler}}, \ and\ \bibinfo {author}
  {\bibfnamefont {A.}~\bibnamefont {Tkatchenko}},\ }\href {\doibase
  10.1103/PhysRevB.86.245405} {\bibfield  {journal} {\bibinfo  {journal} {Phys.
  Rev. B}\ }\textbf {\bibinfo {volume} {86}},\ \bibinfo {pages} {245405}
  (\bibinfo {year} {2012})}\BibitemShut {NoStop}%
\bibitem [{\citenamefont {Ruiz}\ \emph {et~al.}(2016)\citenamefont {Ruiz},
  \citenamefont {Liu},\ and\ \citenamefont {Tkatchenko}}]{PhysRevB.93.035118}%
  \BibitemOpen
  \bibfield  {author} {\bibinfo {author} {\bibfnamefont {V.~G.}\ \bibnamefont
  {Ruiz}}, \bibinfo {author} {\bibfnamefont {W.}~\bibnamefont {Liu}}, \ and\
  \bibinfo {author} {\bibfnamefont {A.}~\bibnamefont {Tkatchenko}},\ }\href
  {\doibase 10.1103/PhysRevB.93.035118} {\bibfield  {journal} {\bibinfo
  {journal} {Phys. Rev. B}\ }\textbf {\bibinfo {volume} {93}},\ \bibinfo
  {pages} {035118} (\bibinfo {year} {2016})}\BibitemShut {NoStop}%
\end{thebibliography}
%

\end{document}